\documentclass[11pt,a4paper]{scrartcl}
\usepackage[utf8]{inputenc}
\usepackage[T1]{fontenc}
\usepackage{lmodern}

\usepackage{graphicx}
\usepackage{epsfig}
\usepackage{float}
\usepackage{cite}
\usepackage{amssymb}
\usepackage{amsmath}
\usepackage{url}
\usepackage[dvipsnames]{xcolor}
\usepackage{cancel}
\usepackage{nicefrac}
\usepackage[normalem]{ulem}
\usepackage[toc,page]{appendix}
\usepackage[onehalfspacing]{setspace}
\usepackage{booktabs}

\usepackage{bigcenter}

\usepackage[format=plain,labelfont={bf},font={small}]{caption}

\usepackage{hyperref}

\usepackage{comment}

\setkomafont{disposition}{\normalcolor\normalfont\bfseries}

\newcommand{\MSbar}{\overline{\text{MS}}}

\newcommand{\refeq}[1]{Eq.~\eqref{#1}}

\newcommand{\reffig}[1]{Fig.~\ref{#1}}

\newcommand{\mr}{\mathrm}

\newcommand{\mur}{\mu_{\text{R}}}
\newcommand{\muf}{\mu_{\text{F}}}
\newcommand{\xir}{\xi_{\text{R}}}
\newcommand{\xif}{\xi_{\text{F}}}

\newcommand{\dsl}{\tilde d_L}
\newcommand{\usl}{\tilde u_L}
\newcommand{\nno}{\tilde \chi_1^0}
\newcommand{\npmo}{\tilde \chi_1^\pm}

\newcommand{\beq}{\begin{equation}}
\newcommand{\eeq}{\end{equation}}
\newcommand{\bea}{\begin{eqnarray}}
\newcommand{\eea}{\end{eqnarray}}

\newcommand{\bec}{\begin{center}}
\newcommand{\eec}{\end{center}}

\newcommand{\becc}{\begin{bigcenter}}
\newcommand{\eecc}{\end{bigcenter}}

\newcommand{\PYTHIA}{\texttt{PYTHIA}}
\newcommand{\PYTHIAE}{\texttt{PYTHIA~8}}

\newcommand{\POWHEGBOX}{\texttt{POWHEG-BOX}}

\newcommand{\Prospino}{\texttt{Prospino}}

\newcommand{\etmiss}{E_T^\mr{miss}}

\newcommand{\ptmiss}{\vec{p\,}_T^\mr{miss}}
\newcommand{\ptjet}{p_T^\mr{jet}}

\newcommand{\utchim}{  \tilde\chi_1^- \tilde u_L}
\newcommand{\utdec}{\tilde u_L \to u \tilde\chi_1^0}
\newcommand{\chidec}{\tilde\chi_1^- \to  e^- \bar\nu_e \tilde\chi_1^0}
\newcommand{\scdec}{e^- \bar\nu_e \tilde\chi_1^0 \, u \tilde\chi_1^0 }

\newcommand{\nnodsl}{pp \rightarrow \nno \dsl  }
\newcommand{\nmousl}{pp \rightarrow \chi_1^- \usl  }

\begin{document}

\renewcommand*{\thefootnote}{\fnsymbol{footnote}}
\begin{flushright}
CERN--TH--2021-–152  \\
PSI-PR-21-23
\end{flushright} 
\begin{center}
	{\Large \textbf{Next-to-leading-order QCD corrections and parton-shower effects for weakino+squark production at the LHC}}\\
	\vspace{.7cm}
	Julien Baglio$^a$\footnote{\texttt{julien.baglio@cern.ch}},
	Gabriele Coniglio$^b$\footnote{\texttt{gabriele.coniglio@uni-tuebingen.de}},
	Barbara J\"ager$^b$\footnote{\texttt{jaeger@itp.uni-tuebingen.de}}, 
	Michael Spira$^c$ \footnote{\texttt{Michael.Spira@psi.ch}}, 
	\\
	\vspace{.3cm}
	$^a$\textit{CERN, Theoretical Physics Department, 1211 Geneva 23, Switzerland\\}
	$^b$\textit{University of T\"ubingen,  Institute for Theoretical Physics, 72076 T\"ubingen, Germany\\}
	$^c$\textit{Paul Scherrer Institute,  5232 Villigen, Switzerland \\}
\end{center}   

\renewcommand*{\thefootnote}{\arabic{footnote}}
\setcounter{footnote}{0}

\vspace*{0.1cm}
\begin{abstract}
	We present a calculation of the next-to-leading order QCD corrections to weakino+squark production processes at hadron colliders and their implementation in the framework of the \POWHEGBOX{}, a tool for the matching of fixed-order perturbative calculations with parton-shower programs. Particular care is taken in the subtraction of on-shell resonances in the real-emission corrections that have to be assigned to production processes of a different type. 
In order to illustrate the capabilities of our code, representative results are shown for selected SUSY parameter points in the pMSSM11. The perturbative stability of the calculation is assessed for the  $\nnodsl$ process. For the squark+chargino production process $\nmousl$ distributions of the chargino's decay products are provided with the help of the decay feature of \PYTHIAE{}.  
\end{abstract}

\tableofcontents

\section{Introduction}\label{s:intro}
%
After the discovery of the Higgs boson~\cite{Aad:2012tfa,Chatrchyan:2012ufa}, with properties compatible with the Standard Model (SM) expectation~\cite{ATLAS:2016neq}, identifying signatures of physics beyond the Standard Model (BSM) remains a prime target of the CERN Large Hadron Collider (LHC). A strong motivation to search for new particles with the properties of Dark Matter~(DM) candidates is provided by astrophysical observations that can best explained in the context of scenarios with massive,  neutral particles that only weakly interact with constituents of the SM. A theoretically appealing class of models to account for such particles is provided by supersymmetric theories that predict a number of new particles differing from their SM counterparts in their spin, and receive larger masses through the soft breaking of supersymmetry (SUSY). The lightest stable SUSY particle (LSP) could constitute a viable DM candidate particle, with all the right properties to account for astrophysical observations. Searching for such particles in the very different environment of a hadron collider provides complementary means to establish the existence of these as of yet hypothetical entities. 

SUSY particles can be produced in various modes at hadron colliders. The largest cross sections are expected for the production of color-charged particles via the strong interaction, e.g.\ QCD-induced pair-production of squarks or gluinos~\cite{Beenakker:1994an, Beenakker:1995fp, Beenakker:1996ch, Beenakker:1997ut, Beenakker:2009ha, Beenakker:2010nq, Beenakker:2011fu, Beenakker:2011sf, Beenakker:2014sma, Beenakker:2016gmf, Beenakker:2016lwe}. However, via their decays into SM particles and lighter, weakly charged SUSY particles squarks and gluinos give rise to rather complicated final states. Alternatively, one could consider the direct production of gaugino pairs~\cite{Beenakker:1999xh}, a gluino in association with a gaugino~\cite{Spira:2002rd} or a squark in association with a neutralino or chargino (here generically referred to as {\em weakinos})~\cite{Binoth:2011xi}. In particular in SUSY scenarios where the lightest neutralino, the $\tilde\chi_1^0$, is the LSP such production modes could result in a pronounced signature with a hard jet (due to the strong decay products of the squarks) and missing transverse energy  stemming from a neutralino~\cite{Binoth:2011xi,Allanach:2010pp}. 

ATLAS and CMS are both exploring final states with one or several hard jet(s) and missing transverse energy as signatures of new physics. For instance, based on data collected in 2015 and 2016 in proton-proton collisions at a center-of-mass energy of  $\sqrt{s}=13$~TeV in final states with an energetic jet and large missing transverse momentum the ATLAS collaboration derived exclusion limits on several models with pair-produced weakly interacting DM candidates, large spatial extra dimensions, and SUSY particles in several compressed scenarios~\cite{Aaboud:2017phn}, where the masses of SUSY particles are very close to each other. 
A similar search by the CMS collaboration resulted in limits on various simplified models of DM and models with large spatial extra dimensions~\cite{Sirunyan:2017jix}. 
The same search strategies could also be applied to interpret data in terms of SUSY scenarios where the lightest neutralino provides a DM candidate and is produced in association with a squark that ultimately gives rise to a hard jet. 

In order to best exploit existing LHC data, accurate predictions for the considered production processes are essential. Ideally, they are provided in the form of Monte-Carlo programs that can easily accommodate experimental selection cuts and be combined with external tools for the simulation of decays, parton-shower, and detector effects. With the work presented in this article, we aim at providing such a program for weakino-squark production processes in the framework of the Minimal Supersymmetric Standard Model (MSSM). The fixed-order perturbative calculation at the heart of our endeavor in many aspects parallels the work of Ref.~\cite{Binoth:2011xi}. However, while that reference focused on providing next-to-leading order (NLO) QCD and SUSY QCD corrections to a wealth of weakino-squark production processes 
and an assessment of multi-jet merging effects at tree level
we provide a fully differential Monte-Carlo program matching the NLO-(SUSY)-QCD calculation with parton showers using the POWHEG method~\cite{Nason:2004rx,Frixione:2007vw} as implemented in the framework of the \POWHEGBOX~\cite{Alioli:2010xd}. Special care is taken in the subtraction of on-shell resonances that might spoil the convergence of the perturbative expansion if not treated properly. To that end, procedures developed and refined in Refs.~\cite{Gavin:2013kga,Baglio:2016rjx,Baglio:2017wka} for the proper treatment of SUSY processes involving on-shell resonances in the context of the  \POWHEGBOX{} are adapted.  
Details of the SUSY parameter spectrum can easily be set via an input file following the SUSY Les Houches Accord (SLHA)\cite{Allanach:2008qq}.

This article is organized as follows: We first describe details of our fixed-order calculation and, in particular, the treatment of on-shell resonances in the framework of the \POWHEGBOX. In Sec.~\ref{s:results} we illustrate the capabilities of our program by applying it to a representative phenomenological study. We select a particular point in SUSY parameter space and consider weakino-squark production at the LHC resulting in signatures with a hard jet and a large amount of missing transverse momentum before concluding in Sec.~\ref{s:concl}. 
We would like to stress that users are free to perform studies of their own design for arbitrary points in the MSSM parameter space with the public version of our code, available from the \POWHEGBOX{} site {\texttt http://powhegbox.mib.infn.it/}.

\section{Details of the calculation}\label{s:calc}
%
\begin{figure}[t]
	\centering
	\begin{tabular}{c c  c}
\includegraphics[height=.2\textwidth]{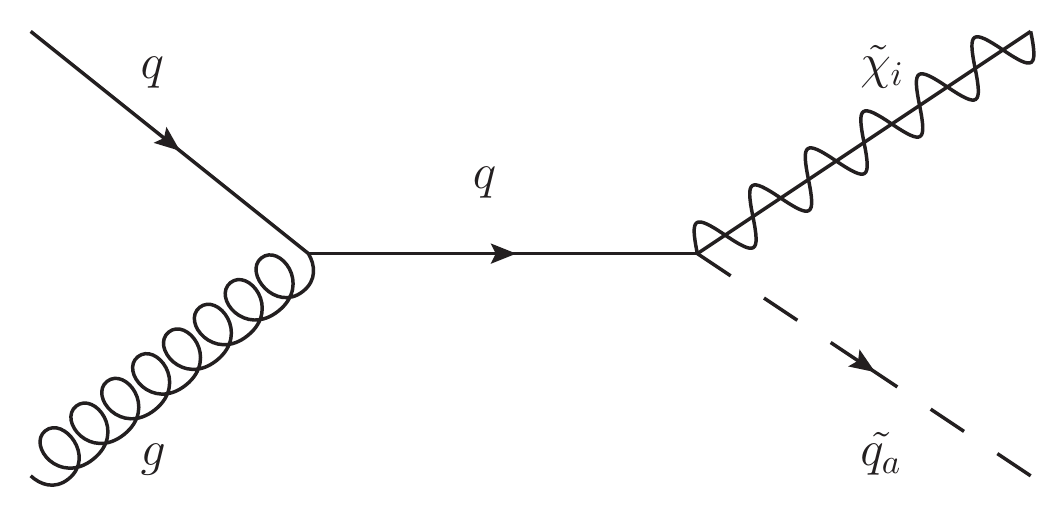} & \hspace{2cm} &
\includegraphics[height=.2\textwidth]{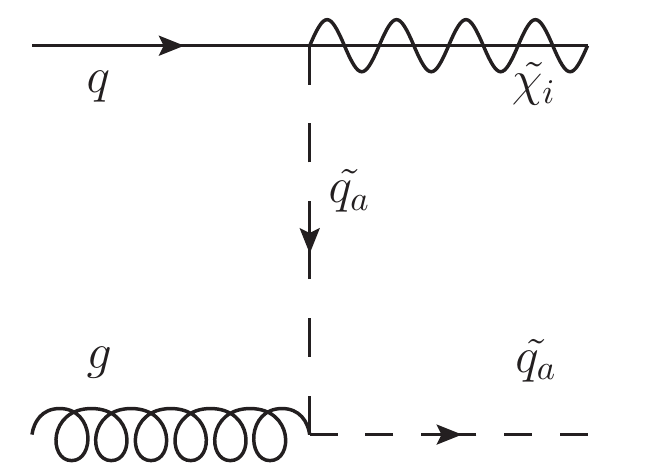}\\
	\end{tabular}
	\caption{Representative tree-level Feynman diagrams for weakino-squark production through an $s$-channel (left) and a $t$-channel(right) process.}
	\label{LO-wesq}
\end{figure}
%
%
Weakino-squark production at hadron colliders at leading order proceeds via two topologies, illustrated in \reffig{LO-wesq}: $s$-channel processes mediated by a quark, and $t$-channel processes mediated by a squark. In both cases, the initial state consists of a gluon and a quark. The weakino in the final state can be either a neutralino or a chargino. A neutralino-squark final state is characterized by a squark that has the same flavor as the quark in the initial state, e.g. $g d  \rightarrow \tilde{\chi}^0_i \tilde{d}_a$, where the index $a=1,2$ represents the chirality index of the squark. The index $i= 1,...,4$ indicates the kind of neutralino produced. A chargino-squark final state will instead exhibit a squark of a different flavor, e.g. $g d \rightarrow \tilde{\chi}^-_i \tilde{u}_a$. In this case, $i=1,2$, as there are only two kind of charginos.  

We consider five massless quarks in the initial state, which allows us to treat the left- and right-handed squarks as mass eigenstates. We consider the CKM matrix to be diagonal.

For our calculation we are considering weakino-squark final states for all types  of squarks and antisquarks of the first two generations and all types of weakinos, leading to a large number of independent channels. For neutralino-squark production, the four flavors of squarks and antisquarks can be produced  both left- and right-handed, leading to 16 production channels for each neutralino and therefore to 64 channels in total, considering the four types of neutralinos. For chargino-squark production, only left-handed squarks and anti-squarks are produced at the considered order of perturbation theory, as the quark-squark-chargino vertex, necessary to have a two-to-two process as seen in \reffig{LO-wesq}, only exists for left-handed squarks. Thus, considering the four flavors of squarks and the two types of charginos and their respective antiparticles, there are 32 channels for chargino-squark production.

The LO amplitudes for all channels have been generated using a tool within the \POWHEGBOX{} suite based on \texttt{MadGraph~4}~\cite{Murayama:1992gi, Stelzer:1994ta, Alwall:2007st}. This tool also provides spin- and color-correlated amplitudes, that are necessary in order to use the automated version of the FKS subtraction algorithm~\cite{Frixione:1995ms} implemented in the \POWHEGBOX{}. To cross-check our results, we verified that these amplitudes were equivalent to those generated using \texttt{FeynArts~3.9}~\cite{Hahn:2000kx} and \texttt{FormCalc~9.4}~\cite{Hahn:1998yk} with the MSSM-CT model file from Ref.~\cite{Fritzsche:2013fta}. 

As a completely alternative approach we used an old (unpublished) calculation performed within the \Prospino{} framework~\cite{Beenakker:1996ed} that relies on a manual generation of all LO, virtual and real correction diagrams that are treated with private implementations of tensor reduction and one-loop scalar integrals. This calculation had been performed with mass-degenerate light-flavor squarks \cite{Beenakker:1994an, Beenakker:1995fp, Beenakker:1996ch, Beenakker:1997ut, Beenakker:1999xh, Spira:2002rd}. We found full numerical agreement for all individual parts in such a mass-degenerate scenario thus corroborating the correctness of the calculation with totally different methods.

%
%
\begin{figure}[tp]
	\centering
	\begin{tabular}{c c c}	
	\textbf{a)} &	\includegraphics[width=.3\textwidth]{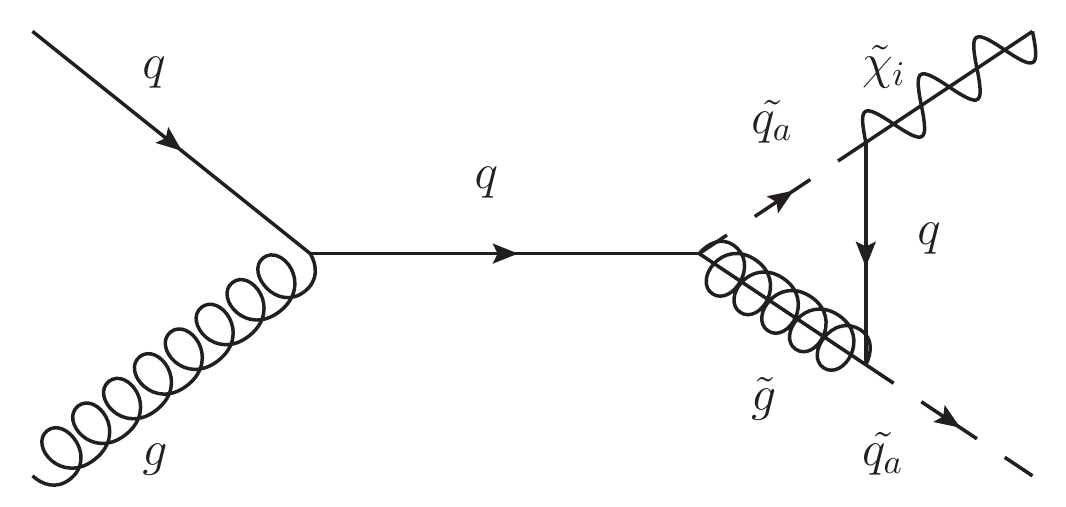} & \includegraphics[width=.3\textwidth]{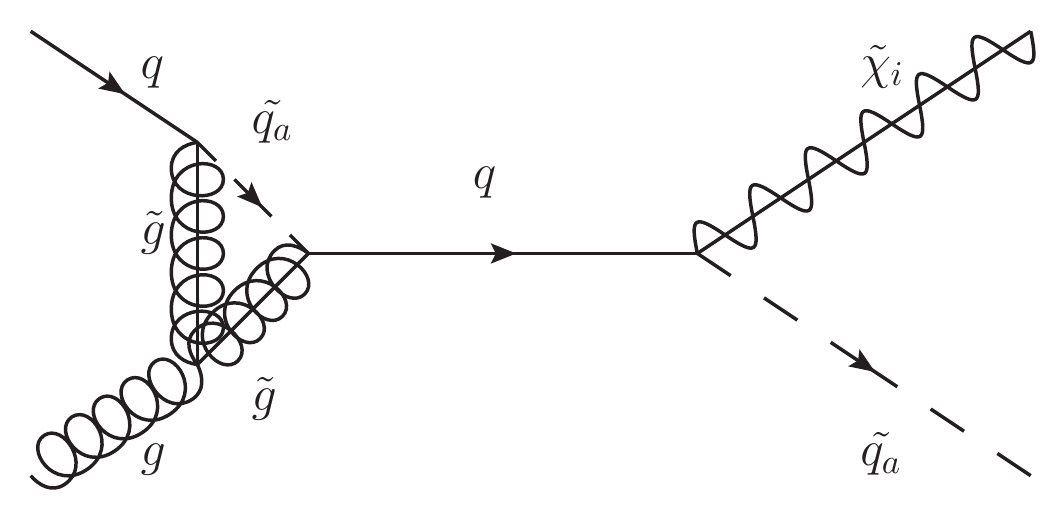} \\
	\textbf{b)} &    \includegraphics[width=.3\textwidth]{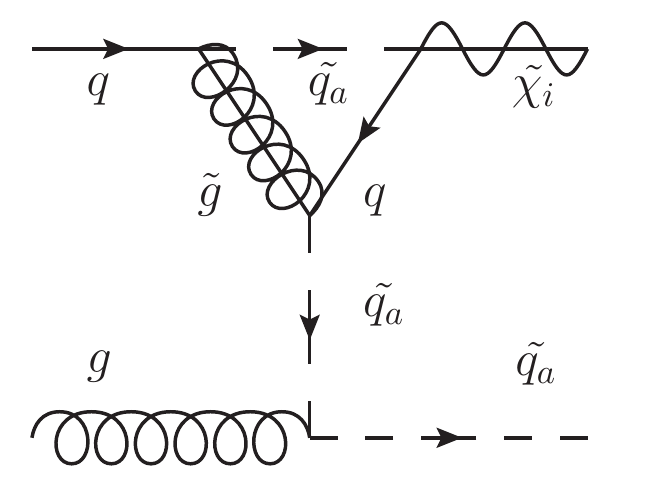} & \includegraphics[width=.3\textwidth]{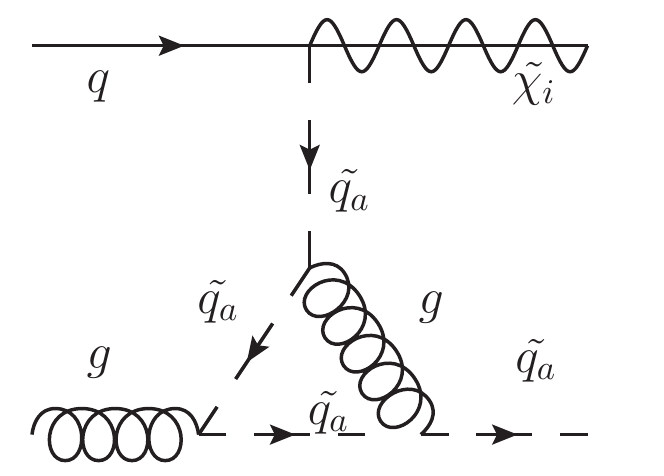} \\
	\textbf{c)} &     \includegraphics[width=.3\textwidth]{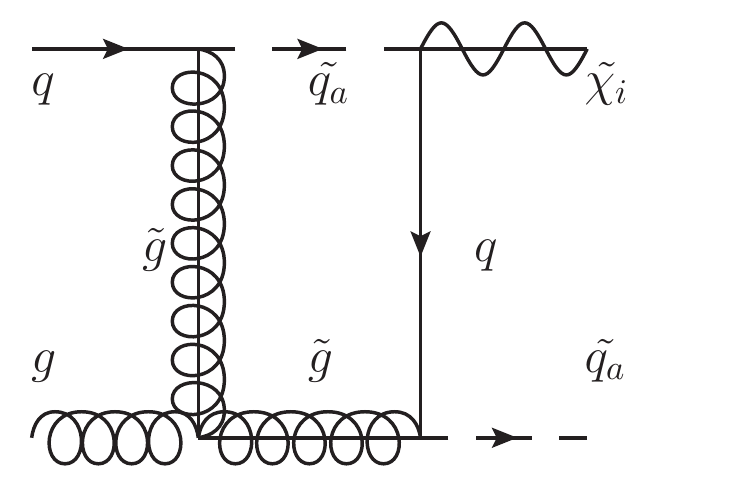} &\includegraphics[width=.3\textwidth]{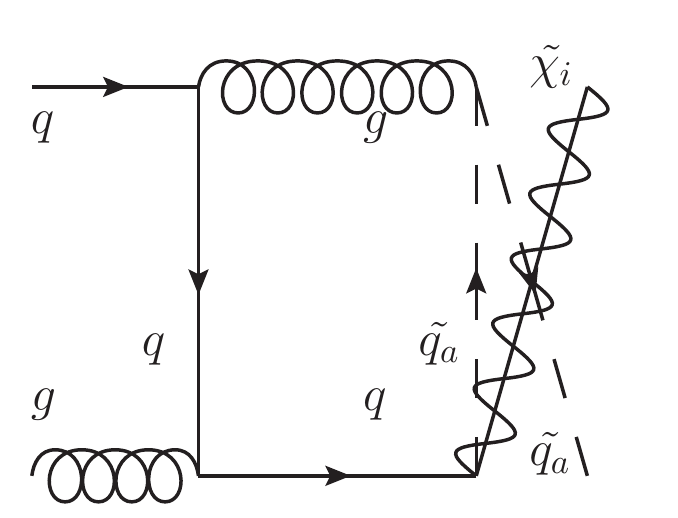}\\
	\textbf{d)} &     \includegraphics[width=.3\textwidth]{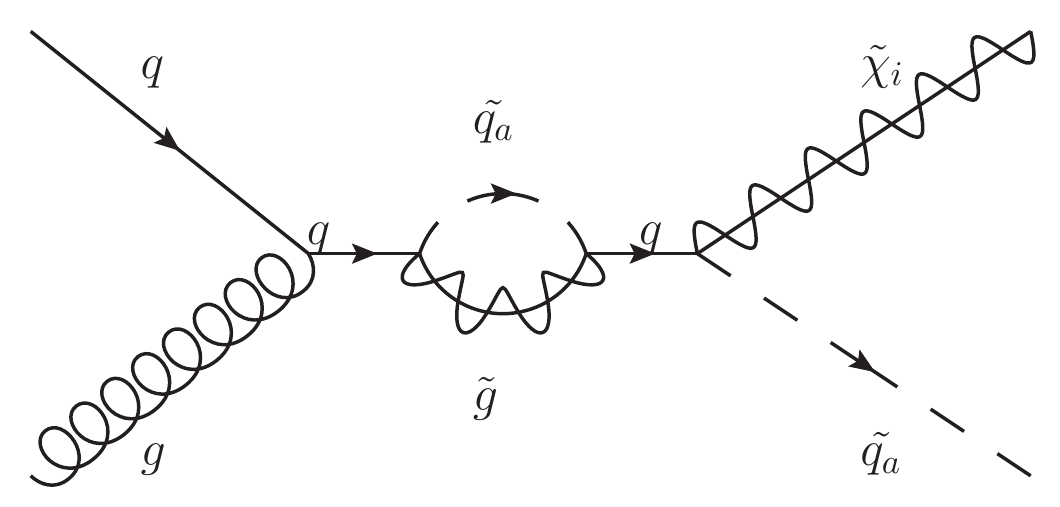} &\includegraphics[width=.3\textwidth]{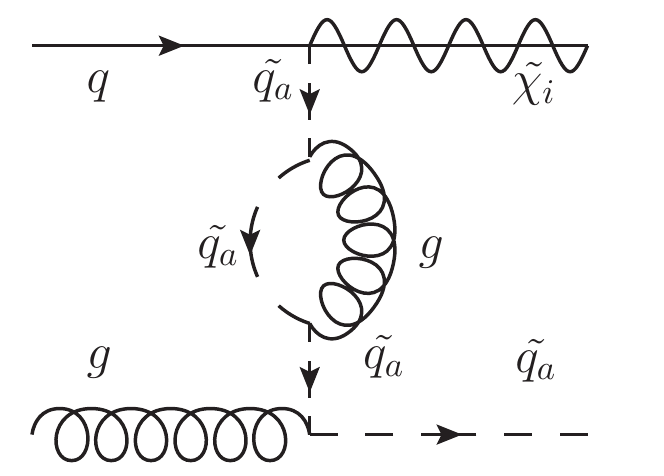}\\
	\end{tabular}
	\caption{Representative Feynman diagrams for the virtual corrections, showing vertex corrections for the $s$-channel diagram (a), vertex corrections for the $t$-channel diagram (b), box corrections (c) and self-energy diagrams (d).}
	\label{virtdiagrams}
\end{figure}
%
%
At next-to-leading order (NLO) in the strong coupling $\alpha_s$, virtual and real-emission corrections have to be computed. 
The virtual corrections include loop corrections to each vertex of the two LO~diagrams and box corrections. There are also self-energy corrections for the intermediate quark of the $s$-channel LO diagram and for the intermediate squark of the $t$-channel LO diagram. Representative diagrams are shown in \reffig{virtdiagrams}. Compared to the LO diagrams, the only new particle appearing in these diagrams is the gluino. The virtual diagrams have been generated using the same tools used to generate the Born ones, \texttt{FeynArts~3.9} and \texttt{FormCalc~9.4}, using the same MSSM-CT model file.

\begin{figure}[t]
	\centering
	\begin{tabular}{c c}
		\includegraphics[width=.3\textwidth]{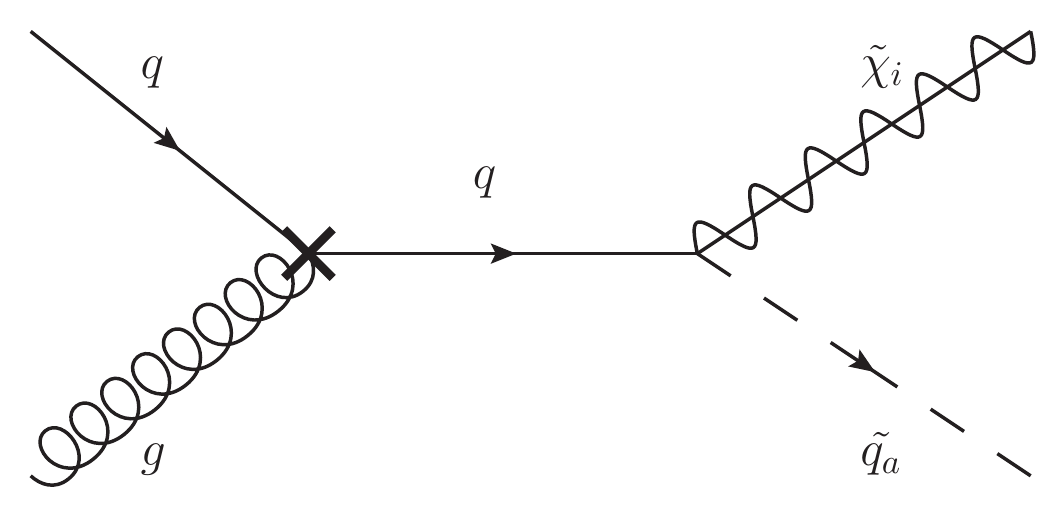} & \includegraphics[width=.3\textwidth]{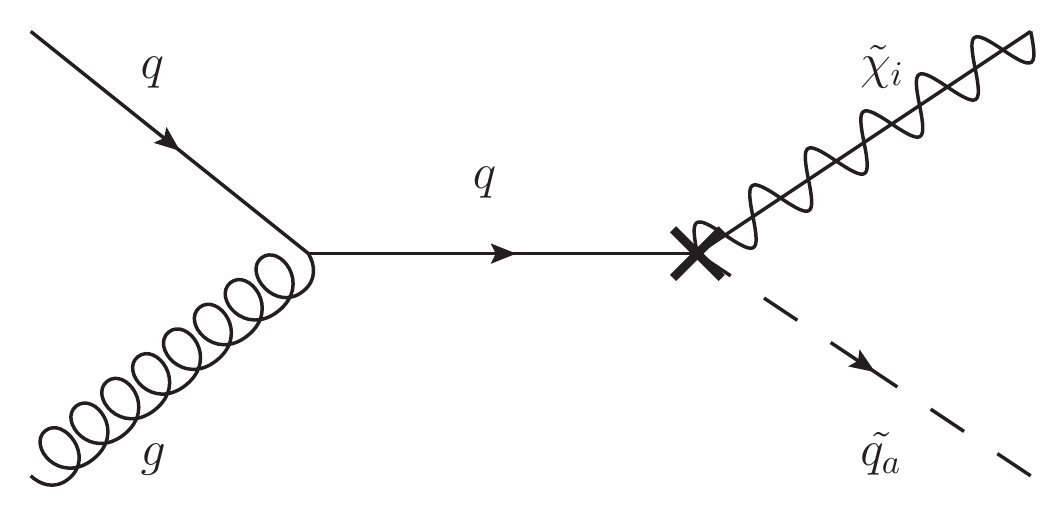} \\
	    \includegraphics[width=.3\textwidth]{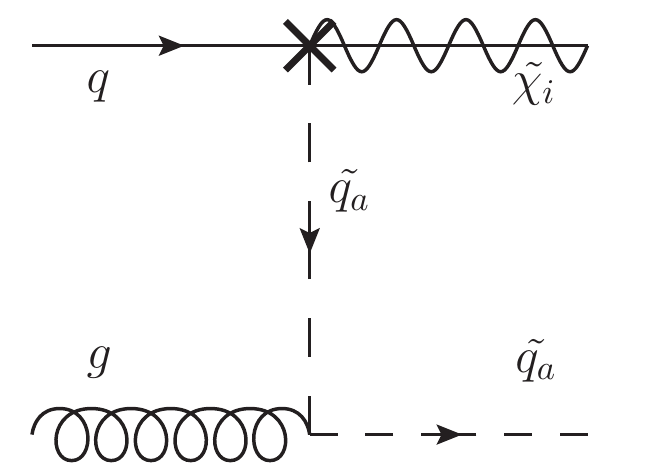} & \includegraphics[width=.3\textwidth]{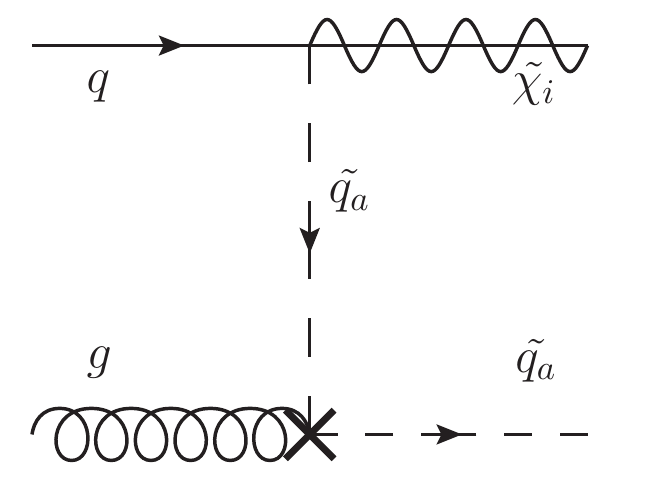} \\
	     \includegraphics[width=.3\textwidth]{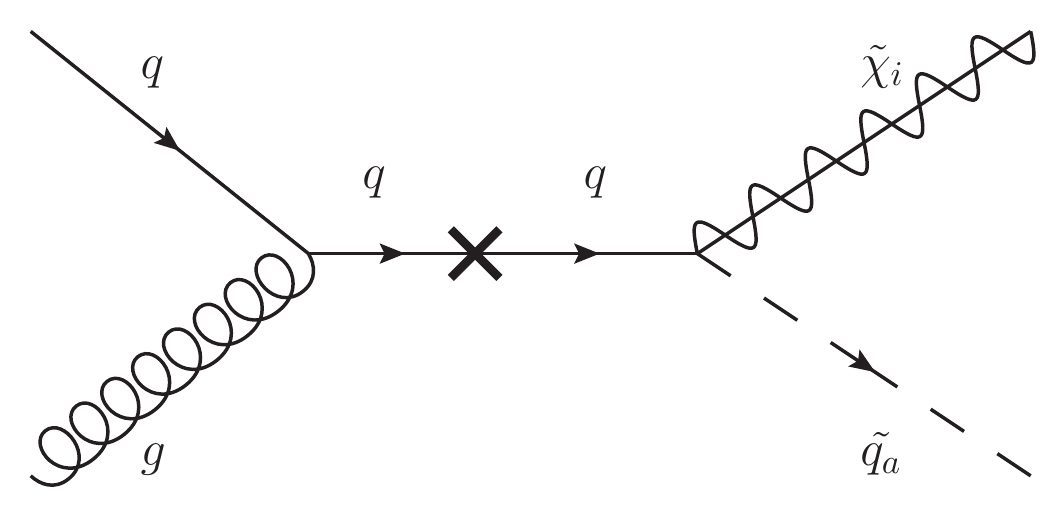} &\includegraphics[width=.3\textwidth]{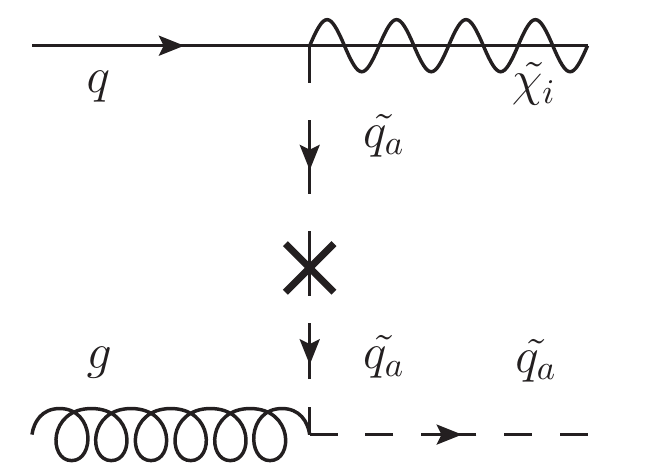}\\
	\end{tabular}
	\caption{Feynman diagrams of vertex (upper and middle rows) and self-energy counterterms (lower row) for weakino-squark production.}
	\label{ctdiagrams}
\end{figure}

Since these diagrams are UV-divergent, a renormalization procedure has to be carried out. The divergences are first regularized using dimensional regularization and then removed using suitable counterterms, which can be generated   by \texttt{FeynArts}.
We use the on-shell scheme to renormalize the wave-functions of the external partons and the masses of the squarks. The strong coupling constant is instead renormalized in the $\MSbar$ scheme with 5 active flavors, i.e.~with decoupled squarks, gluinos and top quark. The Feynman diagrams corresponding to the counterterms, both for the vertices and the self-energy diagrams, are shown in \reffig{ctdiagrams}.

Performing the calculation in $D \neq 4 $ dimensions, as required in the dimensional regularization scheme, strongly breaks supersymmetry introducing a mismatch between the two degrees of freedom of the gaugino and the $(D-2)$ transverse ones of the gauge bosons. This has the practical consequence of spoiling the equality between the gauge couplings and the Yukawa couplings beyond LO, which is required by SUSY invariance. This equality has to be restored introducing finite counterterms \cite{Hollik:2001cz, Martin:1993yx, Beenakker:1999xh}. In our case, the relevant Yukawa coupling is the weakino-quark-squark one, for which the counterterm reads:
\begin{equation}
\hat{g} = g \left( 1 - \frac{\alpha_s}{6 \pi} \right).
\label{susyrestoringCT}
\end{equation}

The virtual diagrams also contain IR divergences, which are ultimately canceled by corresponding divergences  in the real-emission corrections via the FKS subtraction method.

The real emission contributions have been generated using the aforementioned tool based on \texttt{MadGraph~4}. Differently from the LO case with quark-gluon induced channels only, the real corrections include $q \, q $ and $g \, g$ initial states, resulting in the following channels:
\begin{align}
g \, g \rightarrow \tilde{\chi}^i_i \, \tilde{q} \, q, \notag \\
g \, q \rightarrow \tilde{\chi}^i_i \, \tilde{q} \, g, \notag \\
q \, q \rightarrow \tilde{\chi}^i_i \, \tilde{q} \, q. \notag
\end{align} 
%
%
\begin{figure}[t]
	\centering
	\begin{tabular}{c c}
		\includegraphics[width=.3\textwidth]{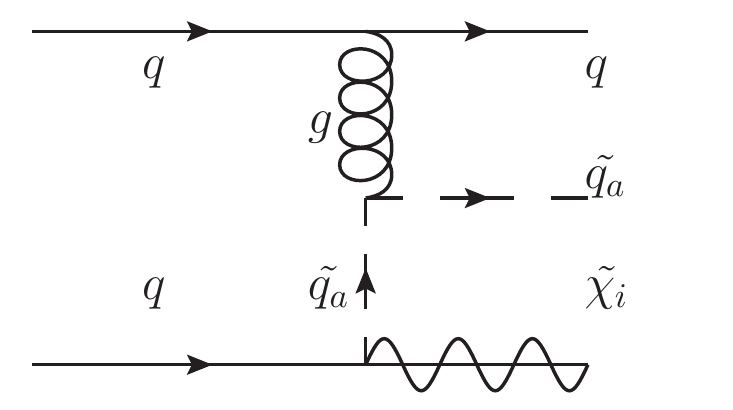} & \includegraphics[width=.3\textwidth]{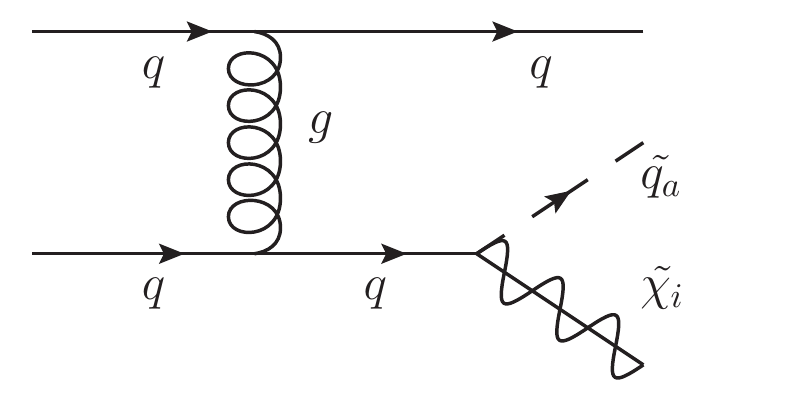} \\
	    \includegraphics[width=.3\textwidth]{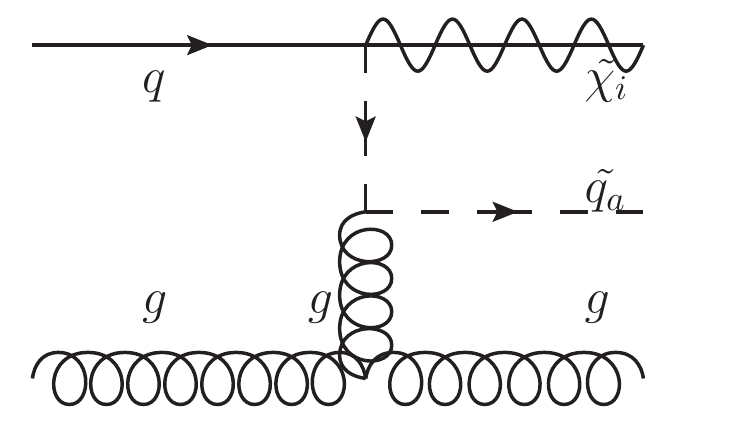} & \includegraphics[width=.3\textwidth]{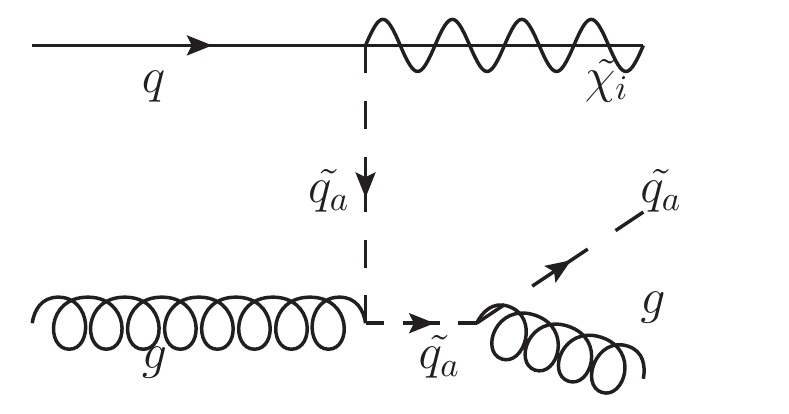} \\
	     \includegraphics[width=.3\textwidth]{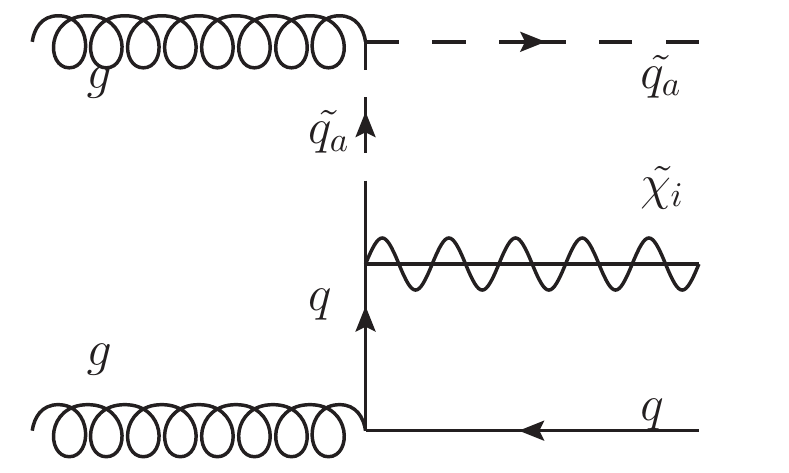} &\includegraphics[width=.3\textwidth]{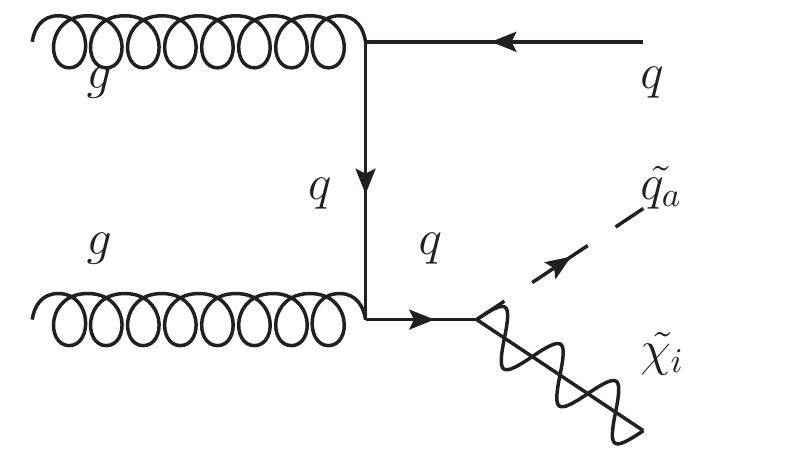}\\
	\end{tabular}
	\caption{Representative Feynman diagrams for the real-emission contributions in the $q \, q$ (upper row), $g \, q$ (central row ) and $g \, g $ (lower row) channels. These diagrams do not include on-shell resonances.}
	\label{nonresreal}
\end{figure}
Representative diagrams are shown in \reffig{nonresreal}. Considering the various possibilities, the real corrections for each final state consist of 30 subchannels for the neutralino-squark final states and of 36 subchannels for the chargino-squark final states.

\begin{figure}[tp]
	\centering
	\begin{tabular}{c c}
		\includegraphics[width=.3\textwidth]{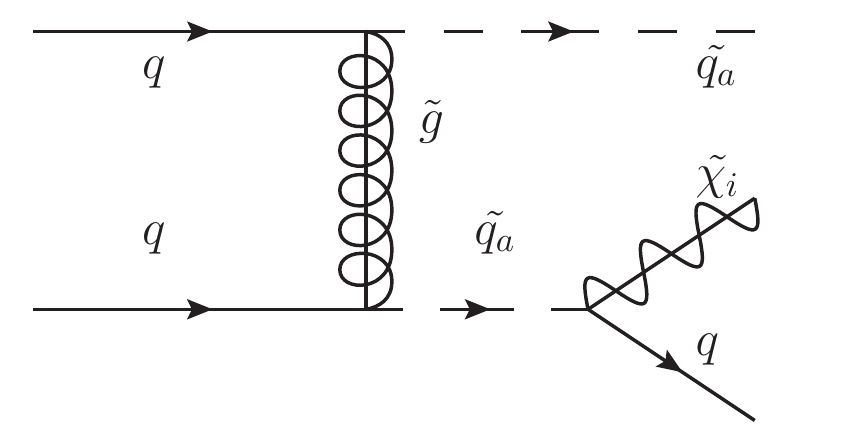} & \includegraphics[width=.3\textwidth]{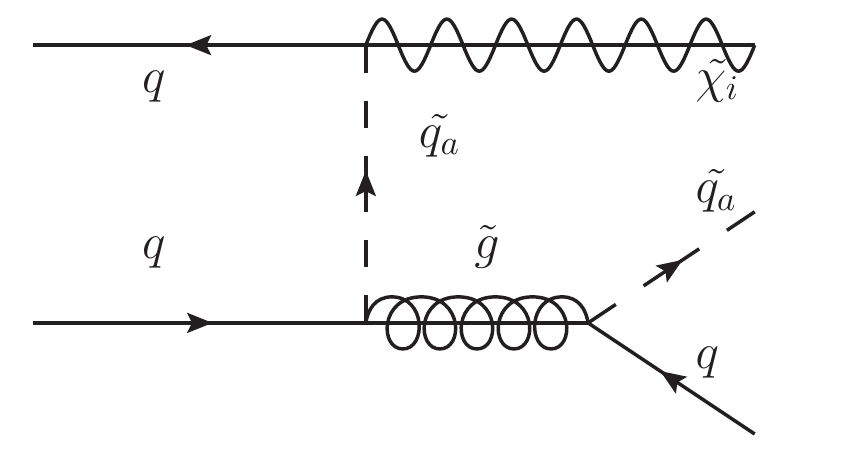} \\
	    \includegraphics[width=.3\textwidth]{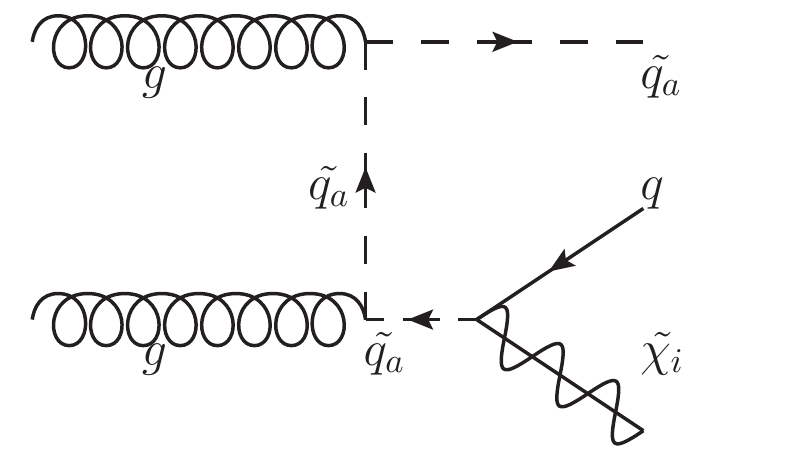} & \includegraphics[width=.3\textwidth]{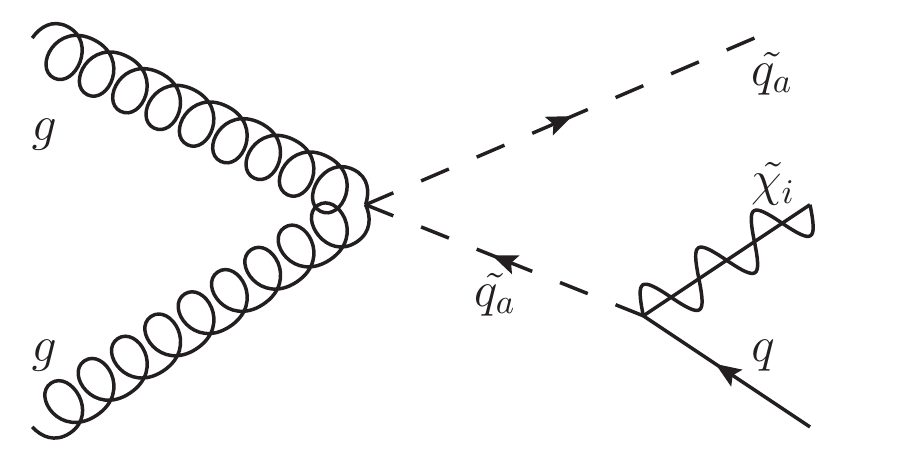} \\
	\end{tabular}
	\caption{Representative Feynman diagrams for real-emission contributions  with on-shell resonances of gluinos (top-right diagram) and squarks (all remaining diagrams). }
	\label{resreal}
\end{figure}

A significant technical complication is represented by the fact that in some of the real-emission diagrams an intermediate particle, namely a squark or a gluino, can become on-shell. Such resonances appear in both the $q \, q$ and the $g \, g$ channels. Representative diagrams are shown in \reffig{resreal}. Squarks of any flavor can become on-shell, as long as they are more massive than the final-state weakino. The on-shell resonances of gluinos are kinematically available when the gluino is heavier than the final-state squark. 
%
%
%
%
Apparently, these resonant contributions spoil the perturbative behavior of our calculation, as they can easily be of the same order of magnitude as the LO cross section. However, these resonant contributions are not a genuine part of the real-emission corrections to the process that we are considering. They should instead be seen as the on-shell production of a different final state followed by a decay into two different particle, e.g.\ in the case of an on-shell resonant squark, as a di-squark production process followed by the decay of one of the squarks into a weakino and a quark. These contributions are therefore already taken into account in the respective production processes. Considering them as a part of our real corrections would lead to them being counted twice. They have therefore to be removed from our real-emission contributions to obtain a well-defined factorization of production and decay processes in the narrow-width approximation, which will then be again perturbatively well-behaved. 

On-shell resonances rarely appear in SM processes (e.g.\  in $W \, t$ production \cite{Re:2010bp}), but are a relatively common feature of SUSY calculations. A similar subtraction procedure is necessary, for instance, in weakino-pair production \cite{Baglio:2016rjx ,Baglio:2017wka} and in squark-pair production \cite{Gavin:2013kga, Gavin:2014yga,Hollik:2012rc}. For the subtraction of on-shell resonances in our calculation we will resort to the procedure discussed in these references\footnote{An alternative treatment is provided by the procedure adopted in the former NLO calculations for squark and gluino \cite{Beenakker:1994an, Beenakker:1995fp, Beenakker:1996ch, Beenakker:1997ut} as well as gaugino pair \cite{Beenakker:1999xh} and associated gaugino-gluino \cite{Spira:2002rd} production where the resonant contribution has been treated with a resonant and factorized parametrization of the phase-space integration within the \Prospino~\cite{Beenakker:1996ed} framework. We have found full agreement between both approaches.}.

The real-emission contributions to a process containing on-shell resonances can be divided in non-resonant (labelled as $nr$) and resonant (labelled as $res$) parts in the following way:
\begin{equation}
|\mathcal{M} |^2 = |\mathcal{M}_{nr} |^2 + 2\, \mr{Re} \left[\mathcal{M}^*_{nr} \mathcal{M}_{res} \right] + |\mathcal{M}_{res} |^2 ,
\end{equation}
where also an interference term between the non-resonant and resonant terms appears. This separation is performed at the diagram level, meaning that only the diagrams that do not contain any resonances are included in the non-resonant matrix elements, $\mathcal{M}_{nr}$. The resonant matrix elements $\mathcal{M}_{res}$ include all the diagrams containing at least one possible on-shell resonance. We have explicitly checked that our matrix element |$\mathcal{M}_{res}|^2$ agrees with an independent derivation starting from the production and decay matrix elements of the intermediate resonant squark/gluino states as has been performed within the \Prospino {} framework in the past. This approach requires the rigorous inclusion of spin correlations and chiral states.

We remove the on-shell contributions from the resonant diagram, performing a pointwise subtraction of a counterterm. 
The first step is the regularization of the singularities present in the propagator of the on-shell particles, achieved by inserting a technical regulator $\Gamma_{reg}$:
\begin{equation}
\label{regprop}
\frac{1}{s_{ij} - m_{ij}^2} \rightarrow \frac{1}{s_{ij} - m_{ij}^2 + i \,  \Gamma_{reg} \, m_{ij} },
\end{equation}
where $m_{ij}$ is the mass of the potentially on-shell particle $ij$ and $s_{ij}= \left( p_i + p_j \right)^2$, with $p_i$ and $p_j$ being the momenta of the final state particles that are daughter particles of the $ij$ particle. The regulator $\Gamma_{reg}$ is not necessarily the physical width of the particle, but a technical parameter. The total cross section, after the removal of on-shell contributions, should not be dependent on its value, as the off-shell contributions are not. This also ensures that, while the applied method in general breaks gauge-invariance, this gauge-invariance breaking effects are removed by going to the narrow-width approximation. This is achieved by approaching the plateau numerically, where the hadronic results do not depend on the regulator $\Gamma_{reg}$ anymore.

After the regularization, the removal of the on-shell contributions is performed by subtracting, locally from each resonant diagram, a counterterm that reproduces the behavior of the on-shell resonances.  For a single on-shell resonance in a $2 \rightarrow 3 $ process the general shape of this counterterm is:

\begin{equation}
|\mathcal{M}_{res}^{\textrm{CT}} |^2 = \prod_{ij} \Theta \left( s - (m_{ij} + m_k^2) \right) \Theta (m_{ij} - m_i - m_j) \, BW \, | \mathcal{M}_{\textrm{res}}|^2_{\textrm{OS}},
\end{equation}
where again the particles originating from the potentially on-shell particle are labelled $i$ and $j$, and the index $k$ denotes the remaining particle, which is often referred to as the spectator particle. The first theta-function represents the condition that the center-of-mass energy squared $s$ is high enough to produce the intermediate particle $ij$ and the spectator particle $k$ on their mass shell. The second theta-function guarantees that the mass of the intermediate particle $m_{ij}$ is larger than the sum of the masses of the two particles $i,j$, as otherwise an on-shell decay would not be possible. The remaining terms are the Breit-Wigner factor, $BW$, and the remapped resonant matrix element squared $|\mathcal{M}_{\textrm{res}}|^2_{\textrm{OS}}$, which is the resonant matrix element squared calculated with on-shell momenta and applying the substitution in \refeq{regprop} to the propagators of on-shell particles. Thus, both terms depend on the regulator $\Gamma_{reg}$.

The Breit-Wigner factor $BW$ is used to suppress the counterterm in correspondence of off-shell regions, to avoid the subtraction of off-shell contributions.  It is defined as the ratio between the matrix element squared and the matrix element squared itself taken in the on-shell limit, i.e.  $s_{ij} \rightarrow m_{ij}^2$ and is equal to: 

\begin{equation}
BW  =\frac{m_{ij}^2 \Gamma^2_{reg}}{(s_{ij}^2 - m^2_{ij})^2 + m_{ij}^2 \Gamma^2_{reg}	}.
\end{equation}
In order for a $BW$ factor to reproduce as closely as possible the behavior of the resonance, the regulator $\Gamma_{reg}$ is usually chosen such that it is as small as possible without causing numerical instabilities in the integration. In the limit $\Gamma_{reg} \rightarrow 0$, the $BW$ factor approaches a Dirac delta, leading to fewer off-shell contributions being included in the counterterm. 

Before integrating the counterterm over the phase space, however, another consideration is necessary: the matrix elements contained in the counterterm have been evaluated in a different phase space that meets the on-shell conditions, which is different from the phase space in which all the other real matrix elements have been calculated, i.e. the general three-particle phase space $d \Phi_3$. Therefore, the integration has to be performed over a separate phase space or, alternatively, a corrective factor reflecting the phase-space transformation, also called Jacobian factor, can be introduced before integrating the real contributions and the counterterms over the same phase space. If $d \tilde{\Phi}_3$ is the on-shell phase space element, it can be expressed in terms of the the regular three-particle phase space as:
\begin{equation}
d \tilde{\Phi}_3 = \mathcal{J}_3 d \Phi_3.
\end{equation} 
The Jacobian factor can be derived by imposing the on-shell condition to the integration over the $d \Phi_3 $ and reads:
\begin{equation}
\mathcal{J}_3 = \frac{s_{ij} \lambda^{1/2} (s, m^2_{ij}, m^2_k) \lambda^{1/2} (m^2_{ij}, m_i^2 , m_j^2) }{m^2_{ij} \lambda^{1/2} (s, s_{ij}, m^2_k) \lambda^{1/2} (s_{ij}, m^2_i, m^2_j)},
\end{equation}
where the Källén $\lambda$ function is defined as $\lambda(x,y,z) = x^2 + y^2 +z^2 - 2xy -2xz -2 yz$.

The cross section for the full real emission contributions can now be calculated as
\begin{equation}
\label{integrreal}
\sigma_{real} = \int d \Phi_3 \left[ |\mathcal{M}_{nr} |^2 + 2 \Re |\mathcal{M}^*_{nr} \mathcal{M}_{res} | \right] + \sigma_{OS},
\end{equation}
where $\sigma_{OS}$ is defined as 
\begin{equation}
\label{integrOS}
\sigma_{OS} = \int d \Phi_3 \left[ |\mathcal{M}_{res} |^2 - \mathcal{J}_3 |\mathcal{M}_{res}^{\textrm{CT}} |^2 \right]\,. 
\end{equation}

There are two possibilities to perform the integration necessary to calculate these cross section. The simpler way is to actually not perform the integration in \refeq{integrOS} before summing it into \refeq{integrreal}, but to perform only one collective integration for the whole real cross section. We will refer to this method as DSUBI. The second possibility is to perform separately the integrations, i.e. summing the contributions from \refeq{integrOS} into \refeq{integrreal} after they have been integrated. We will refer to this method as DSUBII. Requiring two separate integrations, the DSUBII method is obviously more time-consuming. However, it is the default option in our implementation, as for some scenarios, the DSUBI method was leading to numerical instabilities.

To better illustrate the on-shell subtraction scheme we employ let us specifically discuss the real-emission corrections to weakino-squark production. As already mentioned, squarks of every flavor and gluinos can become resonant in this class of processes, leading to 11 independent channels for resonances. In principle, a different regulator could be used for every channel. In our implementation, a distinction is made only between the regulator for squark resonances $\Gamma_{\tilde{q}}$ and the one for gluino resonances $\Gamma_{\tilde{g}}$.

To check the stability of our on-shell subtraction method, we investigated the impact of the variation of the regulator on our results. We noticed that the DSUBI method was sufficient to give stable results when only gluino resonances were present but that when also squark resonances were possible, using the DSUBII method was necessary, because of the more complicated resonant structure. We therefore recommend the use of the DSUBII method. We also observed that the total cross section does not depend on the value of the regulator, as long as this value is chosen to be sufficiently small so that we have numerically reached the narrow-width limit. For both the gluino and the squark resonances, the ratio between the regulator and the mass of the resonant particle should not be larger than $10^{-4}$. Smaller values can be used, but they lead to larger numerical errors.

\begin{figure}[t]
	\centering
	\includegraphics[width=0.7\textwidth]{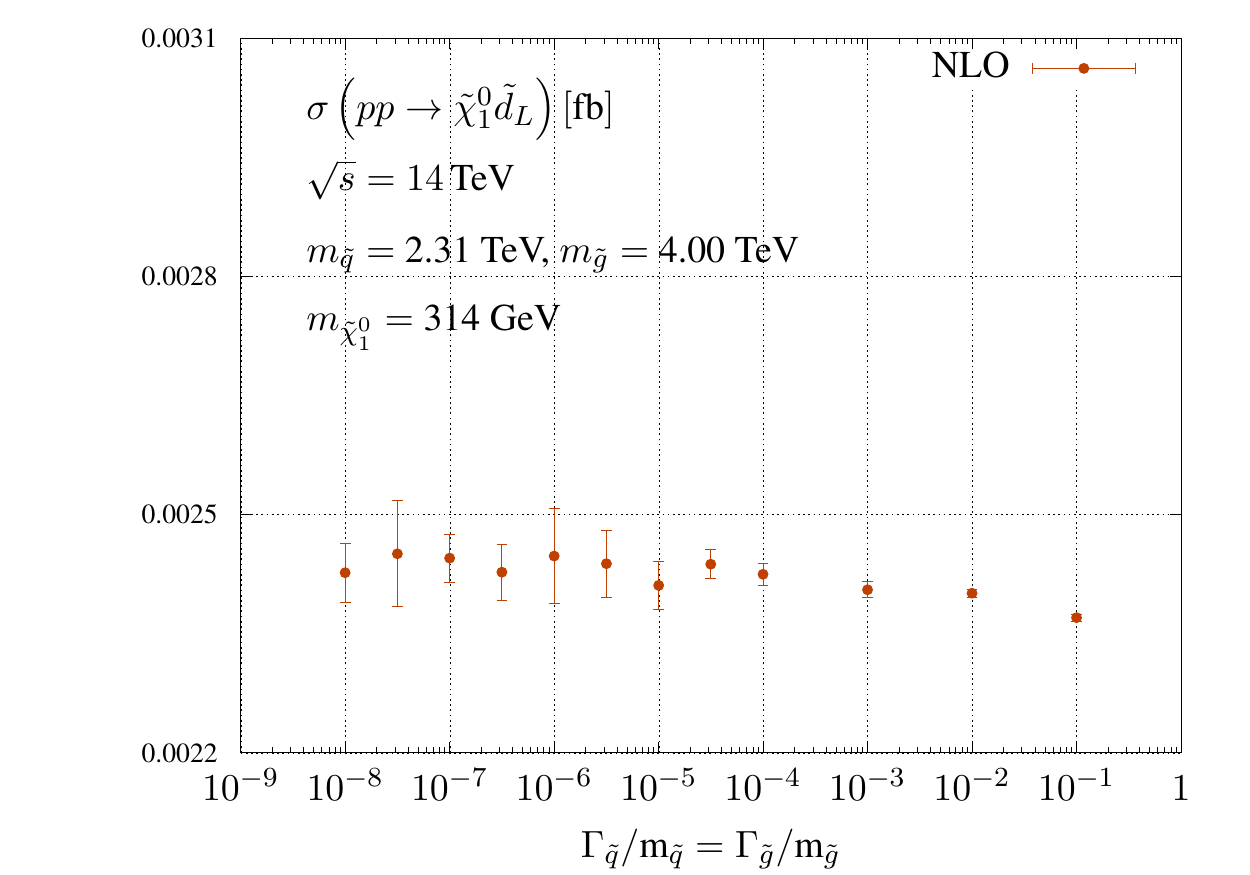} 
	\caption{Dependence of the total cross section for the process $ p p \rightarrow \tilde{d}_L \tilde{\chi}^0_1$ on the regulators  $ \Gamma_{\tilde{q}}$ and $\Gamma_{\tilde{g}}$. The values of the two regulators have been changed simultaneously to explore the range $10^{-8} \leq \frac{\Gamma_i}{m_i} \leq 10^{-4}$, where $i = \tilde{q} , \tilde{g}$.  
	\label{widthdependence}}
\end{figure}

Our findings are shown in \reffig{widthdependence}. We performed the calculation of the NLO cross section for  $\tilde{\chi}^0_1 \tilde{d}_L \,  $ using an artificial SUSY spectrum in which the mass values lead to the simultaneous appearance of gluino and squark resonances. The mass of the neutralino has been set to $314$ GeV, the mass of the gluino has been set to $4.00$ TeV and the squarks have been assumed to be mass degenerate with a mass of $2.31$ TeV. The value of the two regulators has been changed simultaneously to explore the range $10^{-8} \leq \frac{\Gamma_i}{m_i} \leq 10^{-1}$, where $i = \tilde{q} , \tilde{g}$. In the range $10^{-8} \leq \frac{\Gamma_i}{m_i} \leq 10^{-4}$, the cross section appears to be fundamentally independent of the regulator, thus proving that our on-shell subtraction scheme is well-defined and that gauge-violating effects are numerically negligible once the the narrow-width approximation for the intermediate on-shell states has been reached. For larger values of the regulator, the cross section is no longer constant.

As an additional check of our implemention, we reproduced the results presented in Ref.~\cite{Binoth:2011xi} and found, within the attainable accuracy, good agreement with them if adding squarks and antisquarks for the individual production processes.

\section{Phenomenological analysis}\label{s:results}
%
In order to demonstrate the capabilities of our code, we present
numerical results for two  selected scenarios: First, we consider the
on-shell production of a neutralino and a squark for a realistic SUSY
parameter point  in the  phenomenological MSSM model with eleven
parameters, the pMSSM11, suggested in Ref.~\cite{Bagnaschi:2017tru},
which we call {\em scenario~$a$}. This parameter point exhibits both
squark and gluino on-shell resonances, thus showcasing the subtraction
feature of our code. The input parameters and relevant physical masses
are shown in the upper half of Tab.~\ref{SUSYpoints}. Then, we
consider the production of a chargino and a squark for another SUSY
parameter point, extracted from the same reference, which takes into
account constraints from a variety of experiments, including
measurements of the  anomalous magnetic moment of the
muon~\cite{Muong-2:2004fok,Muong-2:2006rrc} (see also the latest
experimental result reported in Ref.~\cite{Muong-2:2021ojo}), which we
call {\em scenario~$b$}. The input parameters and relevant physical
masses are shown in the lower half of Tab.~\ref{SUSYpoints}. 
 
\begin{table}
\begin{center}
\begin{tabular}[b]{| c | c | c | c| c|}
\hline
\multicolumn{5}{|c|}{pMSSM11 - Scenario $a$} \\
\hline
$M_1$ & $M_2$ & $M_3$ & $m_{\tilde{q}}$ &  $ m_{\tilde{q}_3} $ \\

1.3 TeV& 2.3 TeV  & 1.9 TeV & 0.9 TeV & 2.0 TeV \\
\hline
$m_{\tilde{l}}$ & $m_{\tilde{\tau}}$ & $M_A$ & $A$ &  $ \mu $ \\

1.9 TeV & 1.3 TeV & 3.0 TeV & -3.4 TeV & -0.95 TeV \\
\hline
$ \tan \beta $& $m_{\tilde{\chi}^0_1}$ & $m_{\tilde{\chi}^-_1}$ &  $m_{\tilde{d}_L}$ & $m_{\tilde{g}}$  \\

33 & $0.954$ TeV & 0.955 TeV & $1.01$ TeV & $1.98$ TeV \\
\hline
\hline
\multicolumn{5}{|c|}{pMSSM11 - Scenario $b$} \\
\hline
$M_1$ & $M_2$ & $M_3$ & $m_{\tilde{q}}$ &  $ m_{\tilde{q}_3} $ \\

0.25 TeV & 0.25 TeV & -3.86 TeV & 4.0 TeV & 1.7 TeV \\
\hline
$m_{\tilde{l}}$ & $m_{\tilde{\tau}}$ & $M_A$ & $A$ &  $ \mu $ \\

0.35 TeV & 0.46 TeV  & 4.0 TeV & 2.8 TeV & 1.33 TeV \\
\hline
$ \tan \beta $& $m_{\tilde{\chi}^0_1}$ & $m_{\tilde{\chi}^-_1}$ &  $m_{\tilde{u}_L}$ & $m_{\tilde{g}}$  \\

36 & 0.248 TeV & 0.271 TeV & $4.07$ TeV & $3.90$ TeV \\
\hline
\end{tabular}
\end{center}
\caption{Input parameters and relevant physical masses of the SUSY particles in the two scenarios used for our phenomenological studies.}
\label{SUSYpoints}
\end{table}

Our \POWHEGBOX{} code can be used to produce event files in the format of the Les Houches Accord (LHA)~\cite{Alwall:2006yp} for the on-shell production of a squark and a weakino. These event files can in turn be  
processed by a multi-purpose Monte-Carlo program like \PYTHIA{}~\cite{Sjostrand:2006za,Sjostrand:2014zea} that provides a parton shower (PS) to obtain predictions at NLO+PS accuracy. \PYTHIA{} furthermore provides 
the means for the simulation of tree-level decays of unstable SUSY particles. To illustrate that feature, we consider the squark+chargino production channel $pp\to \utchim$  for scenario~$b$  and simulate the decays of the squarks, $\utdec$, and the charginos, $\chidec$ with \PYTHIAE~\cite{Sjostrand:2014zea}, thus providing predictions for $pp\to \scdec$ in the narrow-width approximation for the squark and chargino decays. 
We note that QCD corrections do not affect the purely weak chargino decay. QCD corrections in principle relevant for the squark decay are not taken into account. When labeling the perturbative accuracy of our results, we will only refer to the production process, implicitly assuming that no QCD corrections are provided for the squark decays. 

Throughout our analysis, the renormalization and factorization scales, $\mur$ and $\muf$ are set to 
\beq
\mur=\xir\mu_0\,,
\muf=\xif\mu_0\,,
\eeq
with 
\beq 
\mu_0 = (m_{\dsl} + m_{\chi_1^0})/2\,.
\eeq 
The scale variation parameters $\xir,\xif$ are used to vary the scales around their central value.  If not specified otherwise, they are set to $\xir=\xif=1$.
For the parton distribution functions (PDFs) of the protons we use the CT14LO and CT14NLO set~\cite{Dulat:2015mca} as provided by the LHAPDF library~\cite{Buckley:2014ana} and the associated strong coupling with $\alpha_s(m_Z)=0.118$ for five active flavors. We used as electroweak input parameters the $Z$ boson mass, $m_Z= 91.188$ GeV, the Fermi constant, $G_F= 1.166379\cdot 10^{-5}~\textrm{GeV}^{-2}$ and the $W$ boson mass, $m_W = 80.393$ GeV, while the electromagnetic coupling $\alpha$ is a derived quantity. All our results correspond to proton-proton collisions at a center-of-mass energy of 14~TeV.  Unless stated otherwise, our results do not include contributions from anti-squark production channels.  

\begin{table}
\begin{center}
\begin{tabular}{| c | c | c | c|| c | c | c | c | }
\hline 
 & LO & NLO & $K$ &  & LO & NLO & $K$ \\
\hline
\hline
$ \nno \tilde{d}_L $ & $2.11$  & $3.24$  & 1.54 & $ \nno \tilde{d}_R $ & $0.0348$ & $0.145$ & 4.16   \\
\hline
$ \nno \tilde{u}_L $ & $6.76$  & $9.47$  & 1.42 & $ \nno \tilde{u}_R $ & $0.342 $ & $0.595$ & 1.74  \\
\hline
$ \chi^+_1 \tilde{d}_L $ & $3.80$  & $6.03 $  & $1.59$  & $ \chi^-_1 \tilde{u}_L $ & $9.86$  & $14.2$  & 1.44  \\
\hline
\end{tabular}
\end{center}
\caption{
Cross sections at LO and NLO accuracy and $K$~factors for different final states in scenario~$a$. The quoted numbers include the sum of channels for weakino production in association with a squark and an anti-squark of the given flavor.  All cross-section numbers are given in units of [ab]. The numerical uncertainties do not affect the digits reported. }
\label{xsec table}
\end{table}

In Tab.~\ref{xsec table}, we report cross sections and $K$~factors, defined as the ratio of NLO to LO cross sections,
\beq 
K = \frac{\sigma_{NLO}}{\sigma_{LO}}\,,
\eeq 
for various final states in scenario~$a$. We quote results for the production of squarks and anti-squarks of the first generation in association with either the lightest neutralino~$\nno$ or chargino~$\npmo$. From our results the relevance of the NLO corrections is apparent. An exceptionally large $K$-factor of 4.16 is observed for the $\nno \tilde{d}_R$ channel, due to the suppression of the LO matrix elements and the large numbers of quark- and antiquark-induced channels of the real corrections. 

\begin{figure}[!t]
\bec
\includegraphics[width=0.7\textwidth]{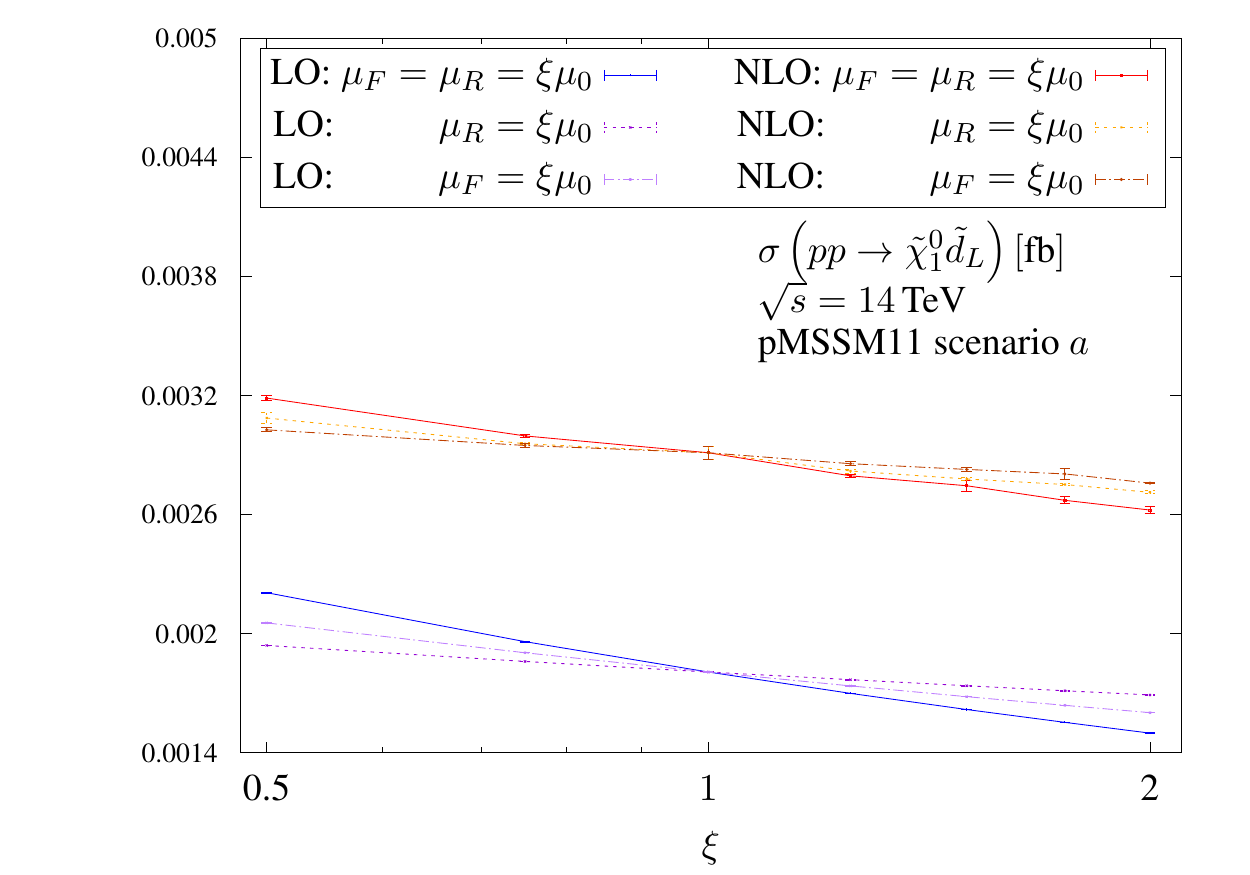}
\caption{Dependence of the total cross section for the process $p p \rightarrow \tilde{d}_L \tilde{\chi}^0_1$ on the renormalization ($\mur$) and factorization($\muf$) scales. Three curves are shown for the LO and NLO cases, respectively. The solid lines correspond to $\mur = \muf = \xi \mu_0$, the dotted lines to $\mur= \xi \mu_0$ with $\muf = \mu_0$ and the dash-dotted lines to  $\muf= \xi \mu_0$ with $\mur = \mu_0$  for scenario~$a$.} 
\label{fig:scaledependence}
\eec
\end{figure}
In~\reffig{fig:scaledependence} we present the dependence of the cross section for the process $p p \rightarrow \tilde{d}_L \tilde{\chi}^0_1$ on the renormalization and factorization scales for scenario~$a$. The cross section depends on both of these scales already at LO. We varied $\mur$ and $\muf$ around the central value $\mu_0= (m_{\tilde{d}_L} + m_{\chi_1^0})/2$ in the range $0.5\mu_0$ to $2\mu_0$. At LO, we observe a more pronounced dependence on the factorization scale,  with the combined variation of factorization and renormalization scale in the above mentioned range leading to a variation of the cross section of approximately $38\%$. At NLO, the dependence on $\xif$ is significantly reduced, while the variation due to $\xir$ is only marginally reduced compared to the LO which can be traced back, to a large amount, to the sizable number of quark- and antiquark-induced channels of the real corrections. The overall variation of the cross section in the range $0.5\mu_0$ to $2\mu_0$ is approximately $19\%$.

For our phenomenological study of the representative squark+neutralino production channel  $pp\to \nno \dsl $  in scenario~$a$ we assume the neutralino gives rise to a signature with a large amount of missing transverse momentum. We do not consider squark decays for this  study, as it is mainly intended to demonstrate the perturbative stability of our results and the applicability of our code to searches for DM  and other types of new physics.   

In this scenario with stable squarks, jets can only arise from real-emission contributions or parton-shower effects. We reconstruct jets from partons using the anti-$k_T$ jet algorithm~\cite{Cacciari:2008gp} with an $R$~parameter of 0.4.  While we do study the properties of such jets to assess the impact of the parton-shower matching on the NLO-QCD results, 
 we do not require the presence of any jets in the event selection, but impose cuts only on the missing transverse momentum of the produced system. 
The missing transverse momentum of an event, $\ptmiss$, is reconstructed from the negative of the vectorial sum of the transverse momenta of all objects assumed to be visible to the detector (i.e.\ jets with a transverse momentum larger than 20~GeV in the pseudorapidity range $|\eta^\mr{jet}|<4.9$ and squarks). The absolute value of $\ptmiss$ is sometimes referred to as ``missing transverse energy'', $\etmiss = |\ptmiss| $. 

%
%
\begin{figure}[t]
\bec
\includegraphics[width=0.52\textwidth]{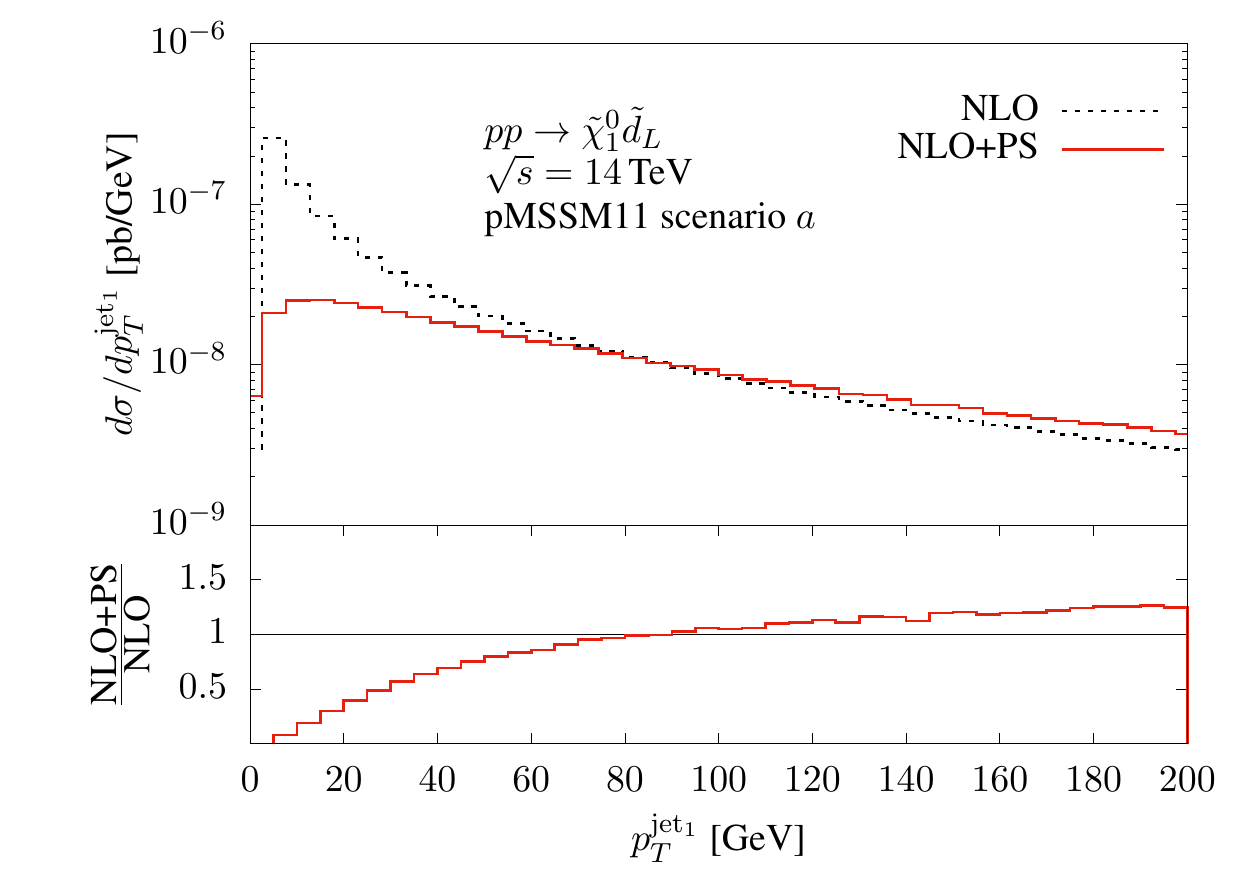}
\caption{Transverse-momentum distribution of the hardest jet in the process $pp\to \dsl \nno$
at NLO (black dotted curve) and NLO+PS (red solid curve) before any cuts are applied. The lower panel shows the ratios of the NLO+PS to the NLO results.  }
\label{fig:ptjet}
\eec
\end{figure}
%
%
In Fig.~\ref{fig:ptjet} we show the transverse-momentum distribution of the hardest jet, before any cuts are imposed. In our setup, with a stable squark, such a jet can only result from the real-emission contributions or from the parton shower. This distribution is thus particularly sensitive to the NLO+PS matching. Indeed, the figure shows the typical Sudakov behavior expected for this distribution: Towards low values of $p_T^\mr{jet}$, the fixed-order result becomes very large. The Sudakov factor supplied by the NLO+PS matching procedure dampens that increase. At higher transverse momenta the NLO+PS results are slightly larger than the fixed-order NLO results. 
Since our analysis is fully inclusive with respect to jets, distributions related to the directly produced squark and neutralino do not exhibit strong sensitivity to this Sudakov damping. 
%

%
%
\begin{figure}[t]
  \becc
\includegraphics[width=0.52\textwidth]{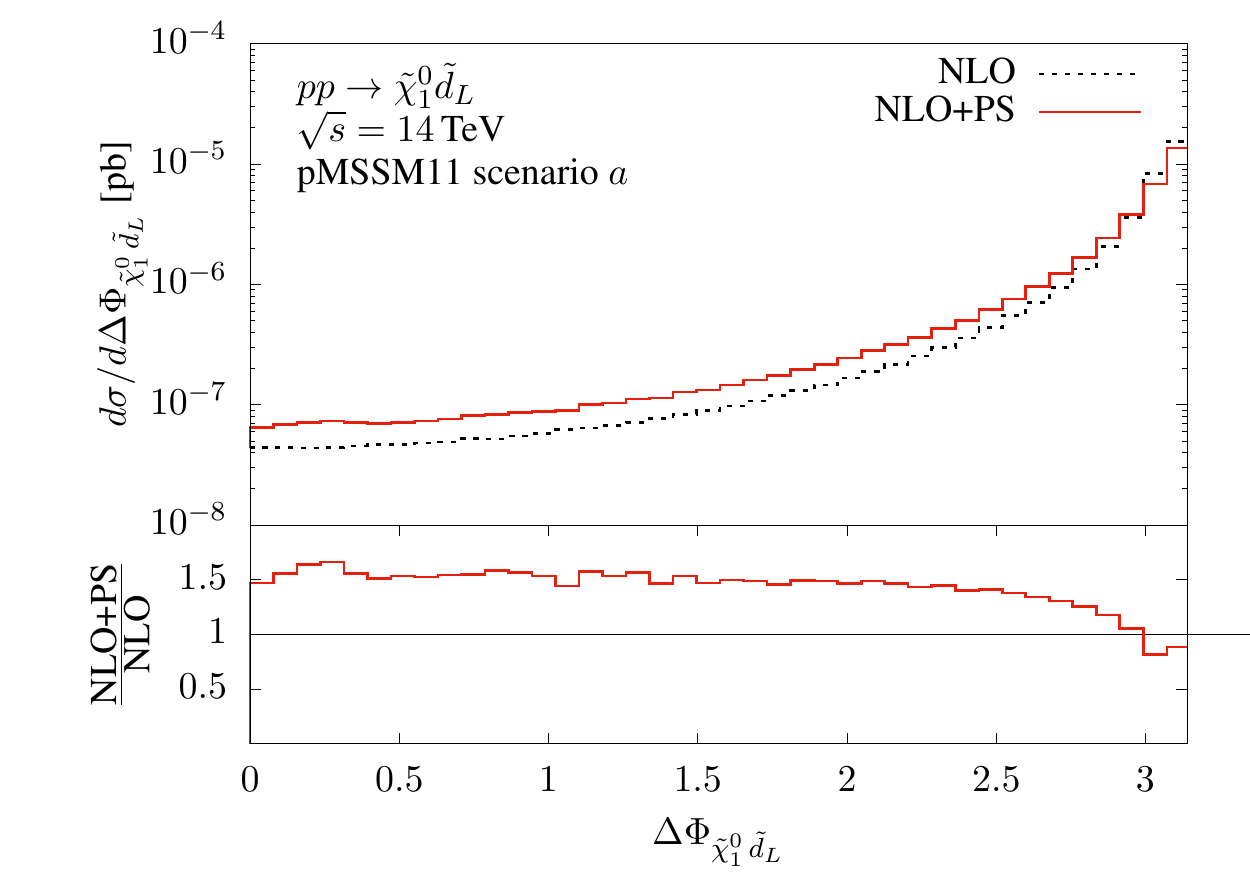}
\includegraphics[width=0.52\textwidth]{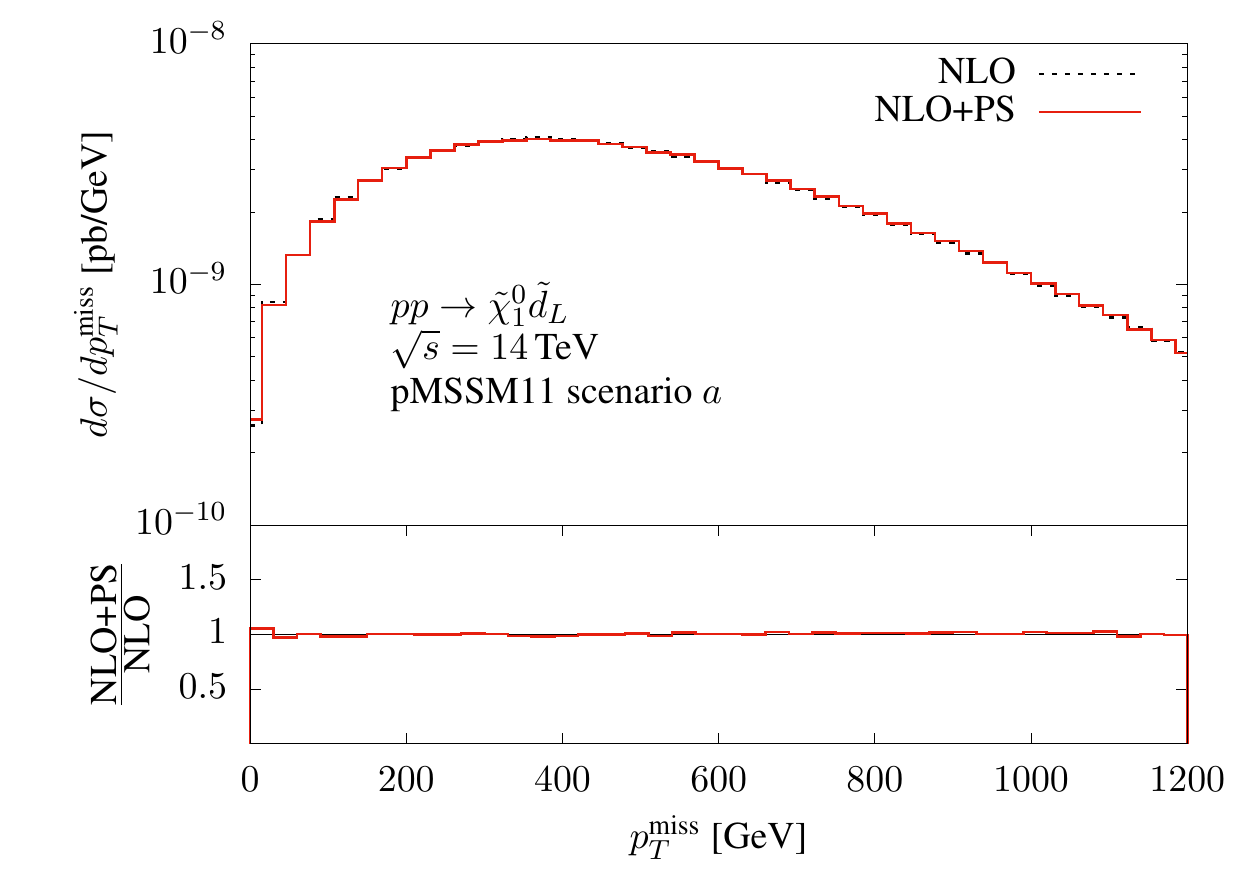}
\caption{Angular separation of the squark and the neutralino (left) and missing transverse energy (right) in the process $pp\to \dsl \nno$
at NLO (black dotted curves) and NLO+PS (red solid curves) before any cuts are applied. The lower panels show the ratios of the respective NLO+PS to the NLO results.}
\label{fig:ptmiss}
\eecc
\end{figure}
%
%
Figure~\ref{fig:ptmiss} illustrates the angular separation of the squark from the neutralino, $\Delta\phi(\nno,\dsl)$,   
and the missing transverse energy at NLO and NLO+PS accuracy before any cuts are imposed. The $\Delta\phi(\nno,\dsl)$ distribution shows that squark and neutralino tend to be produced with large angular separation, and this feature is not significantly altered by parton-shower effects. 
For the missing transverse-momentum distribution, in the bulk the NLO and NLO+PS results are very similar. Only towards very large values of $|\ptmiss|$ where we expect soft radiation effects to become important, the fixed-order NLO result is slightly larger than the corresponding NLO+PS result, which is related to the behavior of the jet at low $\ptjet$ discussed above. Using a cut on the missing transverse energy thus can be considered to be a perturbatively safe choice. 
In the following, in order to consider an event in our analysis, we require it to be characterized by a large amount of missing transverse energy,  
\beq
\label{eq:cut-ptmiss}
|\ptmiss|>250~\mr{GeV}\,.  
\eeq

%
%
\begin{figure}[t]
\becc
\includegraphics[width=0.52\textwidth]{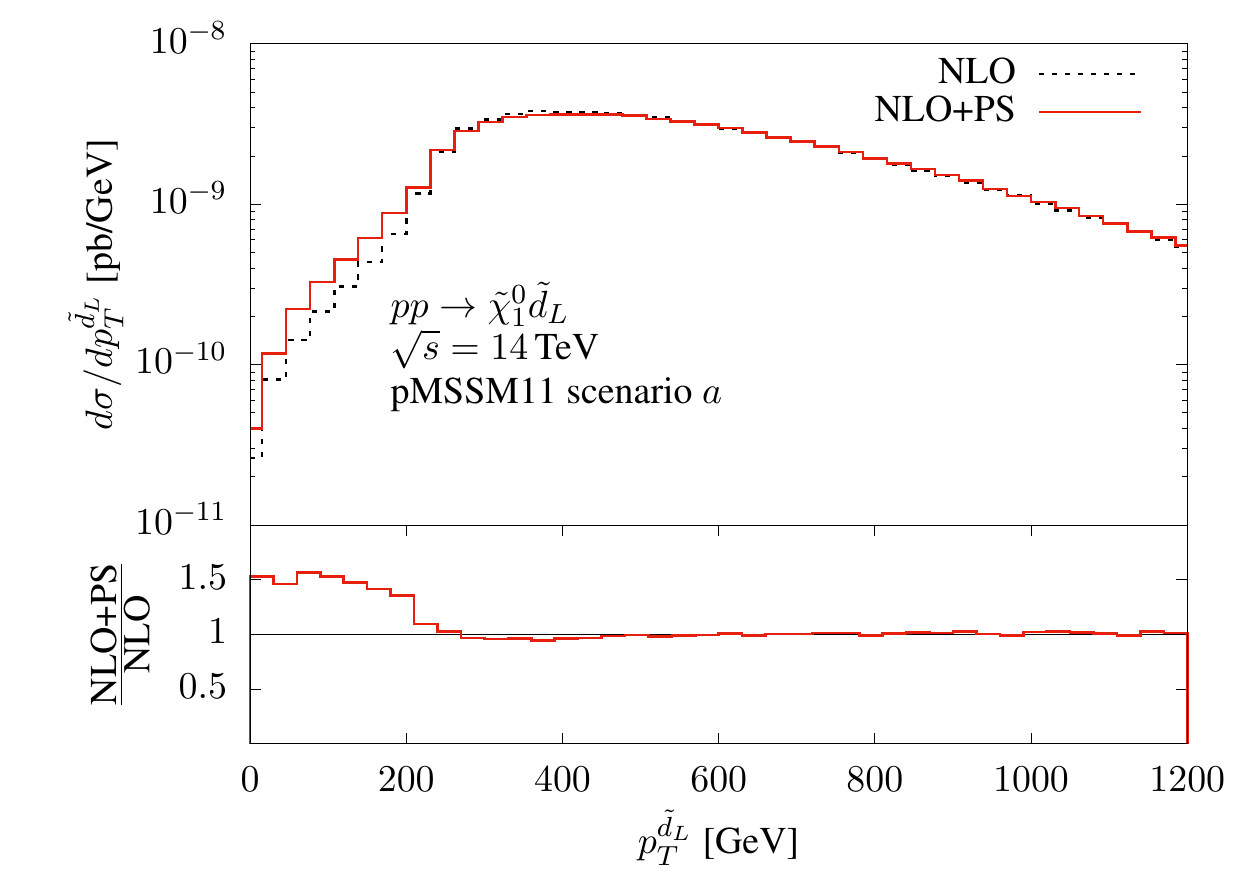} 
\includegraphics[width=0.52\textwidth]{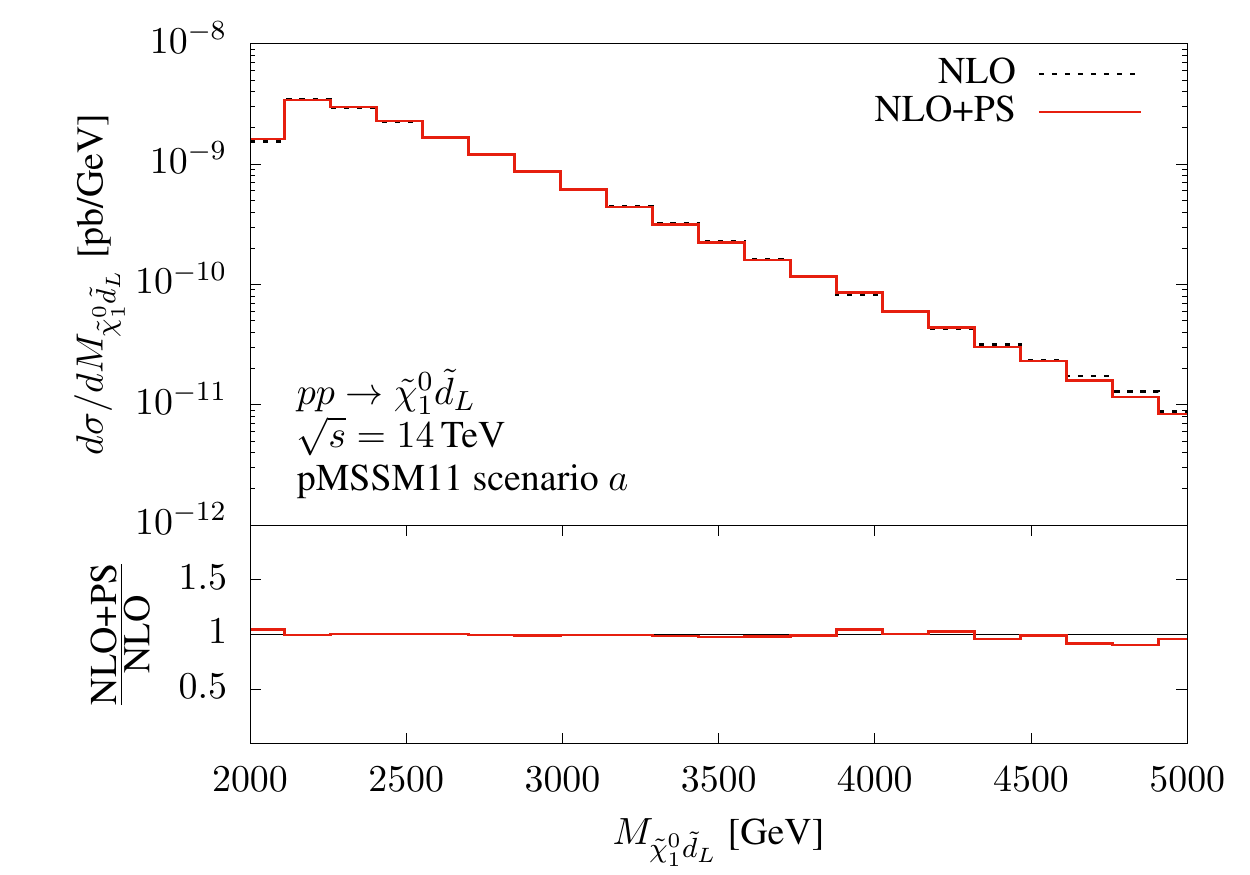}
\caption{Transverse momentum  of the squark (left) and invariant mass of the squark-neutralino system (right) in the process $pp\to \dsl \nno$
at NLO (black dotted curves) and NLO+PS (red solid curves), after the cut of Eq.~(\ref{eq:cut-ptmiss}) has been applied. The lower panels show the ratios of the respective NLO+PS to the NLO results.}
\label{fig:ptsquark}
\eecc
\end{figure}
%
%
In Fig.~\ref{fig:ptsquark} we show the transverse-momentum distribution of the squark after the cut of Eq.~(\ref {eq:cut-ptmiss}) has been imposed. Slight differences between the NLO and the NLO+PS predictions occur only in the region of low transverse momenta, where soft-gluon effects of the Sudakov factor dominate the NLO+PS results. Even smaller differences occur in the invariant mass distribution of the  squark-neutralino system, depicted in the same figure. 

\begin{figure}[t]
\becc
\includegraphics[width=0.52\textwidth]{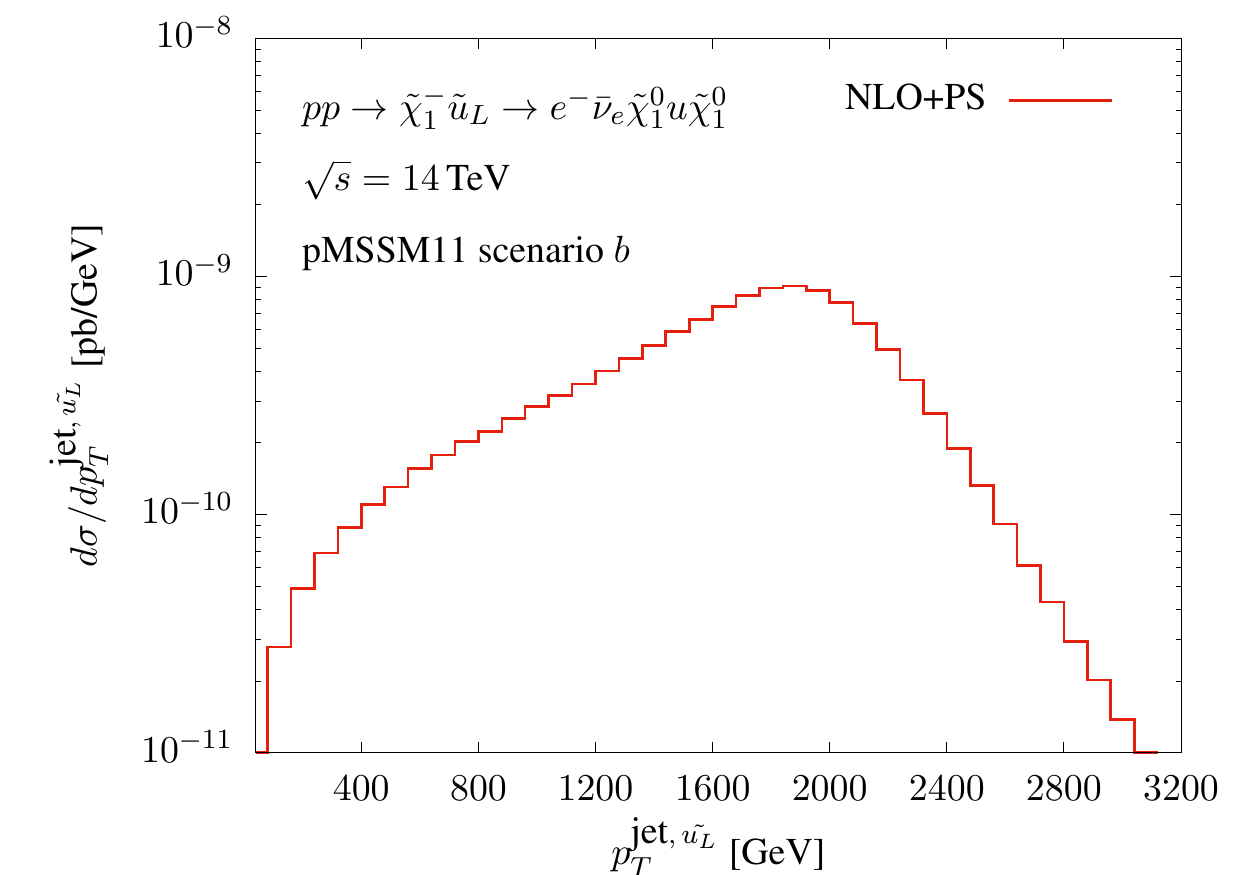}
\includegraphics[width=0.52\textwidth]{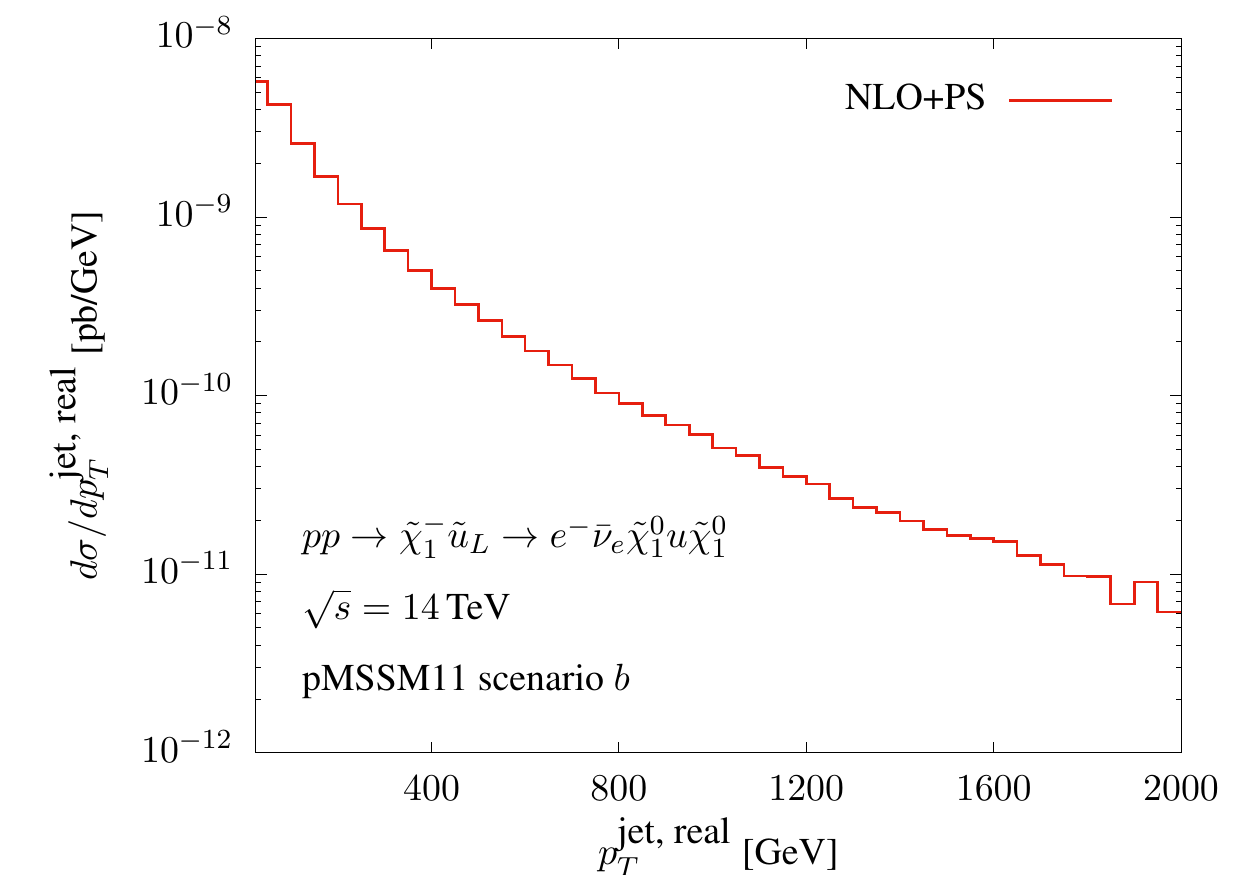}
\caption{Transverse-momentum distribution of the jet stemming from the squark decay (left) and of the real-emission parton (right) in the process 
$pp\to \utchim\to \scdec$ 
at NLO+PS accuracy before any cuts are applied. }
\label{fig:chargino-ptjet}
\eecc
\end{figure}

Next, let us consider the squark+chargino production channel $pp\to \utchim$ at NLO+PS accuracy, combined with tree-level decays of the squark, $\utdec$, and the chargino, $\chidec$,  as provided by \PYTHIAE. 
For the considered parameter point, the quark resulting from the squark gives rise to a very hard jet, shown in Fig.~\ref{fig:chargino-ptjet}. 
%
%
%
%
Further jets can be generated by real-emission corrections and the parton-shower.  As demonstrated by Fig.~\ref{fig:chargino-ptjet}~(r.h.s), such jets exhibit an entirely different shape. 

\begin{figure}[!h]
\bec
\includegraphics[width=0.52\textwidth]{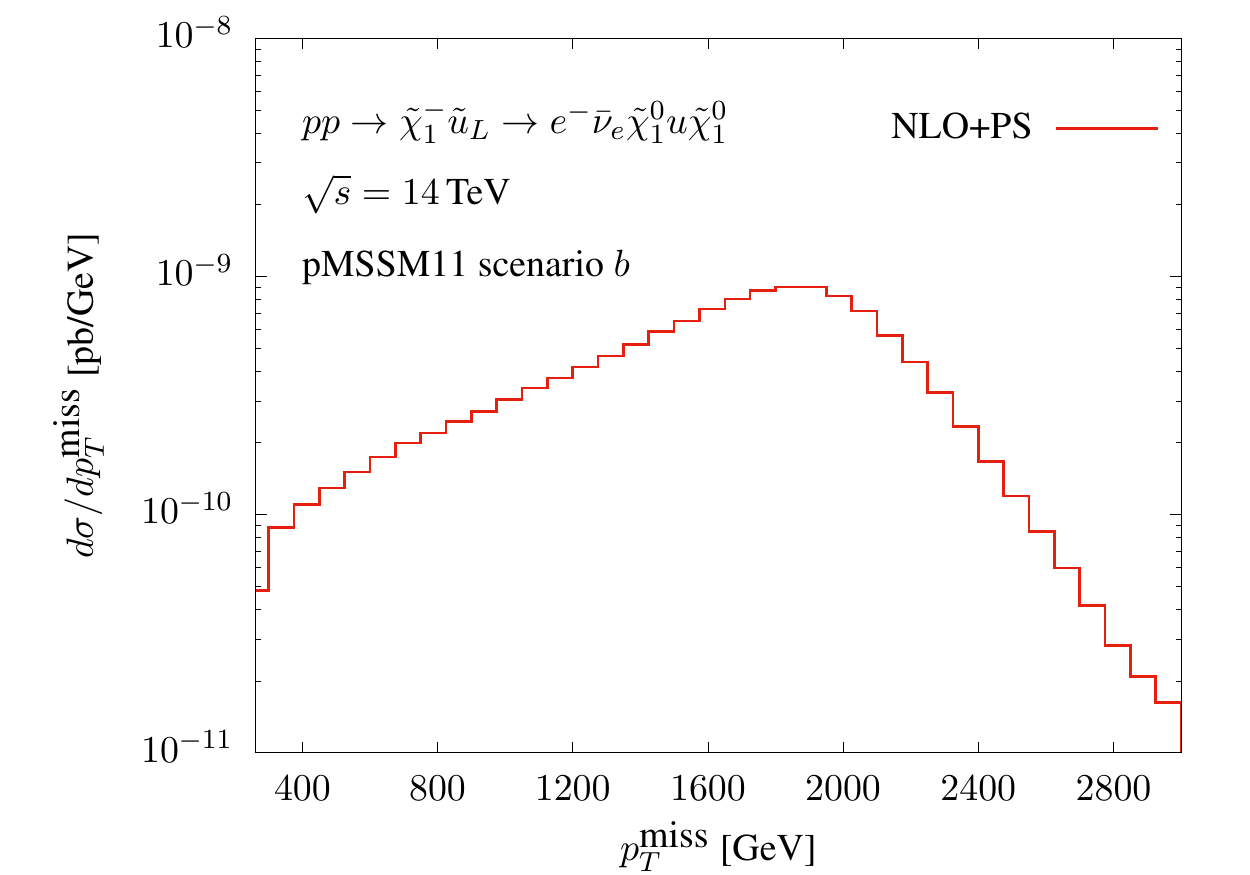}
\caption{Missing transverse energy of the final-state system in the process $pp\to \utchim\to \scdec$
at NLO+PS accuracy before any cuts are applied.}
\label{fig:chargino-etmiss}
\eec
\end{figure}

In addition to the jet, the squark decay gives rise to a very hard neutralino. It contributes to the missing momentum of the final-state system together with the neutralino and the neutrino stemming from the chargino decay and jets that are too soft to be identified. The total missing transverse energy of the $\scdec$ final state is depicted in Fig.~\ref{fig:chargino-etmiss}. 
%
%
%
%
From this distribution it is clear that we can afford to impose a hard cut on $\etmiss$ to suppress background processes with typically much smaller amounts of missing transverse momentum,  without significantly reducing the signal cross section. In the following we therefore impose a cut of 
\beq
\label{eq:chargino-cut}
\etmiss \geq 250~\mr{GeV}\,, 
\eeq
which reduces the cross section for $\scdec$ production by less than one percent. 

\begin{figure}[!h]
\becc
\includegraphics[width=0.52\textwidth]{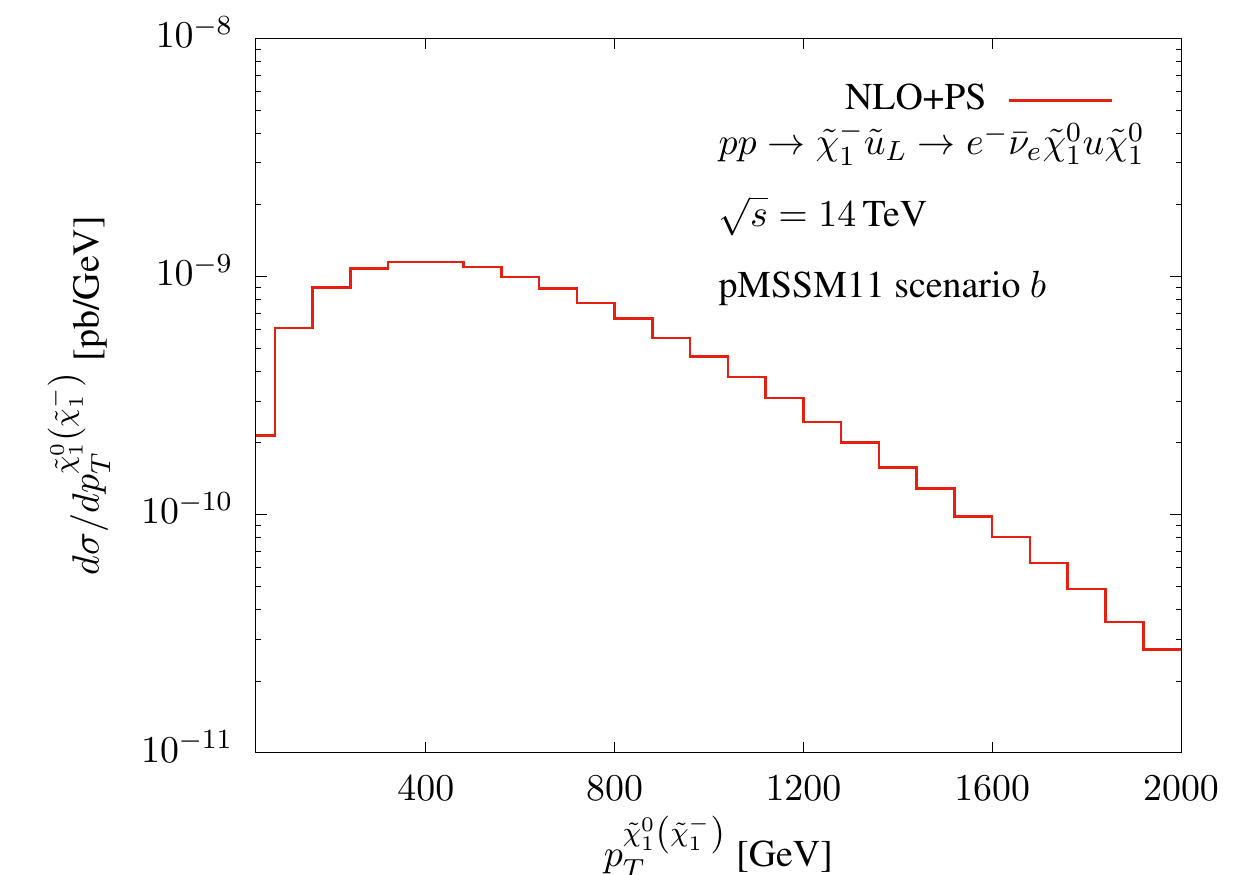}
\includegraphics[width=0.52\textwidth]{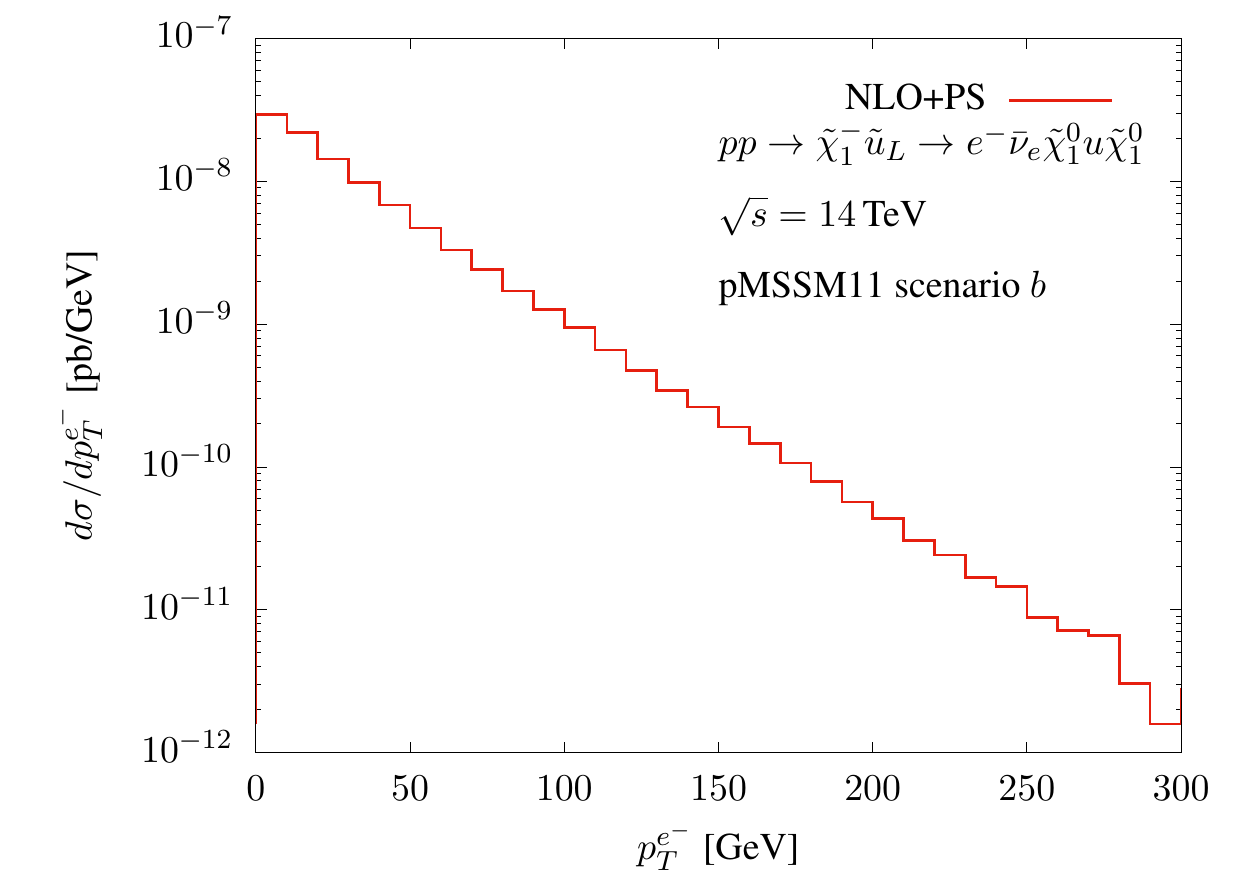}
\caption{Transverse momentum distribution of the neutralino (left) and the electron (right) stemming from the chargino decay in the process 
$pp\to \utchim\to \scdec$ 
at NLO+PS accuracy, after the cut of Eq.~(\ref{eq:chargino-cut}) has been applied. }
\label{fig:chargino-decay}
\eecc
\end{figure}

Because of the mass pattern of mother and decay particles, the momentum balance of the chargino-decay system is quite uneven: The heavy neutralino acquires a large amount of transverse momentum, peaked at about 450~GeV, while the lepton exhibits a much softer distribution, see Fig.~\ref{fig:chargino-decay}. 
%
%
%
Even softer leptons are expected in SUSY scenarios where the masses of the chargino and its decay neutralino are yet closer than for the considered pMSSM11 point. This feature has to be taken into account in searches aiming to use the lepton to tag a particular signal. 

\section{Conclusions}\label{s:concl}
%
In this article we presented a calculation of the NLO-QCD corrections to the entire range of weakino+squark production processes, and their matching to parton-shower programs. We discussed in detail the subtraction of on-shell resonances that appear in the real-emission corrections, but de facto should be considered as part of a different production process. Two completely independent implementations of the fixed-order calculation were devised and compared to ensure the reliability of our work. One of these constitutes the central element of a new \POWHEGBOX{} implementation that will be made public at \href{http://powhegbox.mib.infn.it/}{http://powhegbox.mib.infn.it/}. 

In order to illustrate the capabilities of this implementation, we presented results for two parameter points of the pMSSM11 for the representative $pp \to \nno \dsl $ and $pp\to \utchim$ processes, considering tree-level decays of the squark and chargino in the latter. We found that NLO-QCD corrections are of a significant size, increasing the LO estimate by approximately~50\%.  The additional effect of the parton shower on NLO predictions is in general moderate, with the exception of regions in phase space where resummation effects become important. 
With the help of a multi-purpose Monte-Carlo generator like \PYTHIA{} decays of the unstable SUSY particles of the squark+weakino production process can be simulated.  

We wish to point out that in this article we only highlighted some representative applications of the program we developed, and hope that henceforth the tool will find ample use in customized applications by the phenomenological and experimental high-energy communities.

\section*{Acknowledgements}
The authors would like to thank Wim Beenakker, Michael Kr\"amer, Tilman Plehn and Peter Zerwas for valuable discussions and their collaboration in unpublished, but closely related work on these processes for mass-degenerate squark flavors in the \Prospino{} framework. The authors would also like to thank Christoph Borschensky and Matthias Kesenheimer for valuable discussions. Part of this work was performed on the high-performance computing resource bwForCluster NEMO with support by the state of Baden-W\"urttemberg through bwHPC and the German Research Foundation (DFG) through grant no INST 39/963-1 FUGG. 

\clearpage

\providecommand{\href}[2]{#2}\begingroup\raggedright\endgroup

\end{document}